\documentclass[12pt,nofootinbib,superscriptaddress]{revtex4}

\usepackage{amssymb,color,paralist,amsmath,epsfig}
\usepackage{graphicx}
\usepackage{ }
\usepackage{amsfonts}

\usepackage{relsize}

\usepackage[colorlinks,citecolor=blue,linktoc=all,linkcolor=cyan]{hyperref}

\newcommand{\beq}{\begin{equation}}
\newcommand{\eeq}{\end{equation}}

\newcommand{\ber}{\begin{eqnarray}} 
\newcommand{\eer}{\end{eqnarray}}

\begin{document}

\title{Two-photon exchange correction to muon-proton elastic scattering at low momentum transfer}

\author{Oleksandr Tomalak}
\affiliation{Institut f\"ur Kernphysik, Johannes Gutenberg Universit\"at, Mainz, Germany}
\affiliation{PRISMA Cluster of Excellence, Johannes Gutenberg-Universit\"at,  Mainz, Germany}
\affiliation{Department of Physics, Taras Shevchenko National University of Kyiv, Ukraine}
\author{Marc Vanderhaeghen}
\affiliation{Institut f\"ur Kernphysik, Johannes Gutenberg Universit\"at, Mainz, Germany}
\affiliation{PRISMA Cluster of Excellence, Johannes Gutenberg-Universit\"at,  Mainz, Germany}

\date{\today}

\begin{abstract}
We evaluate the two-photon exchange (TPE) correction to the muon-proton elastic scattering at small momentum transfer.  Besides the elastic (nucleon) intermediate state contribution, which is calculated exactly, we account for the inelastic intermediate states by expressing the TPE process approximately through the forward doubly virtual Compton scattering. The input in our evaluation is given by the unpolarized proton structure functions and by one subtraction function. For the latter, we provide an explicit evaluation based on a Regge fit of high-energy proton structure function data. It is found that,  
for the kinematics of the forthcoming muon-proton elastic scattering data of the MUSE experiment, the elastic TPE contribution dominates, and the size of the inelastic TPE contributions is within the anticipated error of the forthcoming data. 
\end{abstract}

\maketitle

\tableofcontents

\section{Introduction}
\label{sec1}

The present measurements of the proton charge radius from muonic hydrogen spectroscopy \cite{Pohl:2010zza, Antognini:1900ns} differ by a puzzling $7 \sigma$ from the radius value extracted from the hydrogen spectroscopy \cite{Mohr:2012tt} and the elastic electron-proton scattering data \cite{Bernauer:2013tpr}. This huge discrepancy, which become known as the "proton radius puzzle", see Ref. \cite{Carlson:2015jba} for a recent review, 
has led to intense theoretical and experimental activity in recent years. So far it has defied an explanation which could bring all three experimental techniques in agreement with each other.  
To shed a further light on this puzzle, several new experiments involving muons are being planned. Their aim is to test the lepton universality in the interaction of a lepton with a proton. One can compare the elastic scattering of electrons and muons on the proton target and measure the proton charge radius in the muon-proton elastic scattering in a similar way as it was done in the electron-proton elastic scattering \cite{Rosenbluth:1950yq,Bernauer:2013tpr}. Such an elastic scattering experiment is presently being planned by the MUSE Collaboration~\cite{Gilman:2013eiv}. 
Complementary, one can also compare the electron and muon pair photoproduction on the proton as proposed in \cite{Pauk:2015oaa}. These experiments should be performed at the $ 1 ~\% $ level or better of experimental accuracy in order to have an impact on the observed discrepancy in the proton charge radius. Such a level of precision therefore calls for studies of the higher order corrections in such processes, as the corrections to cross sections suppressed by one power in the fine-structure constant $\alpha = e^2/(4 \pi) \approx 1/137$ are also in the 1 \% or few \% range. 
In particular it requires studies of the two-photon exchange (TPE) correction to the unpolarized lepton-proton elastic scattering for the case when the mass of the lepton cannot be neglected relative to its momentum.  
In a previous work~\cite{Tomalak:2014dja}, we have performed an estimate of the leading TPE contribution from the proton intermediate state, and provided estimates for the MUSE experiment. In this work we account for the inelastic intermediate states, i.e. all possible intermediate states in the TPE box graphs beyond the proton state. 
As our aim is an estimate of such corrections for the MUSE experiment, which corresponds with very low momentum transfers, 
we will estimate the inelastic TPE corrections through the near forward doubly virtual Compton scattering process \cite{Tomalak:2015aoa}. The essential hadronic information is contained in the unpolarized proton structure functions and in one subtraction function.  We clarify the TPE correction coming from the subtraction function, and provide an empirical  determination of the subtraction function based on the high-energy behavior of the forward Compton amplitude $ \mathrm{T}_1 $.

The plan of the present paper is as follows. 
We review the elastic TPE contribution to the unpolarized lepton-proton scattering in Sec. \ref{sec2}, and derive a low-momentum transfer expansion accounting for all terms due to the non-zero lepton mass.  We subsequently discuss the forward unpolarized doubly virtual Compton scattering process which will serve as our starting point in the determination of the inelastic TPE corrections in Sec. \ref{sec3}. We evaluate the TPE correction due to the subtraction function in the forward Compton amplitude  $ \mathrm{T}_1 $, and provide an empirical determination of the subtraction function from data in Sec. \ref{sec4}. We provide the expressions for the inelastic TPE correction coming from the unpolarized proton structure functions and present the results of our numerical evaluation 
for the muon-proton elastic scattering in Sec. \ref{sec5}.  Our conclusions are given in Sec. \ref{sec6}.

\section{Elastic TPE contribution}
\label{sec2}

\begin{figure}[h]
\centering{\includegraphics[width=.35\textwidth]{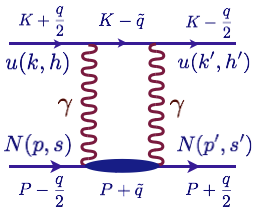}}
\caption{TPE graph in elastic lepton-proton scattering.}
\label{TPE_kinematics}
\end{figure}

In this work we study the TPE correction at low momentum transfer to unpolarized elastic scattering of a  charged lepton with mass $ m $ and initial (final) momentum $ k $ ($ k' $) on a proton with mass $ M $ and initial (final) momentum $ p $ ($ p' $), see Fig. \ref{TPE_kinematics} for the notations of kinematics and helicities.  In this work, we consider the region of small squared momentum transfer $ Q^2 \ll ~M^2,~M E $, where $ E $ is the lepton beam energy (in the lab frame) and $ Q^2 = -\left(k-k'\right)^2 $.

We define the TPE correction $ \delta_{2 \gamma} $ from the difference between the expected cross section $ \sigma^{\mathrm{exp}} $ and the cross-section in the one-photon exchange (OPE) approximation $ \sigma_{1 \gamma} $:
\ber \label{TPE_definition}
\sigma^{\mathrm{exp}} \simeq \sigma_{1 \gamma} \left(1 + \delta_{2 \gamma} \right),
\eer
corresponding with first order corrections in the fine structure constant $ \alpha = e^2 / \left( 4 \pi \right) $, with $ e $ the unit of electric charge.

In the low $ Q^2 $ region, the dominant contribution to the TPE graph of Fig. \ref{TPE_kinematics} results from the proton intermediate state (elastic contribution). We have fully calculated this contribution in a previous work \cite{Tomalak:2014dja}. In this section, we start by studying the quality of some approximate expressions for the elastic contribution in the low $Q^2$ limit, for which analytical expressions can be provided.

The first TPE estimate is due to Feshbach and McKinley \cite{McKinley:1948zz}, who calculated the TPE contribution, corresponding with Coulomb photon couplings to the static proton (i. e. two $ \gamma^0 $ vertices). This so-called Feshbach term contribution to $ \delta_{2 \gamma} $ in Eq. (\ref{TPE_definition}) is denoted by $ \delta_F $ and can be expressed through the scattering angle in the laboratory frame $ \theta $ and the lepton velocity $ v $ as
\ber \label{full_feshbach}
\delta_F = \pi \alpha v  \frac{ \sin \theta/2 \left( 1 -  \sin \theta/2 \right) }{1 - v^2 \sin^2 \theta/2 }.
\eer

As a next step, one may consider the TPE correction in the scattering of two point-like Dirac particles (corresponding with two $ \gamma^\mu $ couplings). In Appendix \ref{lowQ2limit} we provide some analytical expressions of this contribution in the limit of small $ Q^2 \ll M^2, ~M^2 \left( E^2 - m^2 \right)/s $, with the center-of-mass frame squared energy $ s = M^2 + m^2 +2 M E $, both for the cases when $ Q^2 \ll m^2 $, see Eqs. (\ref{TPE_expansion_massive}-\ref{IR_piece}), and when $ Q^2 $ and $ m^2 $ are of similar size, see Eq. (\ref{elastic_TPE_general}).

\begin{figure}[h]
\centering{\includegraphics[width=1.\textwidth]{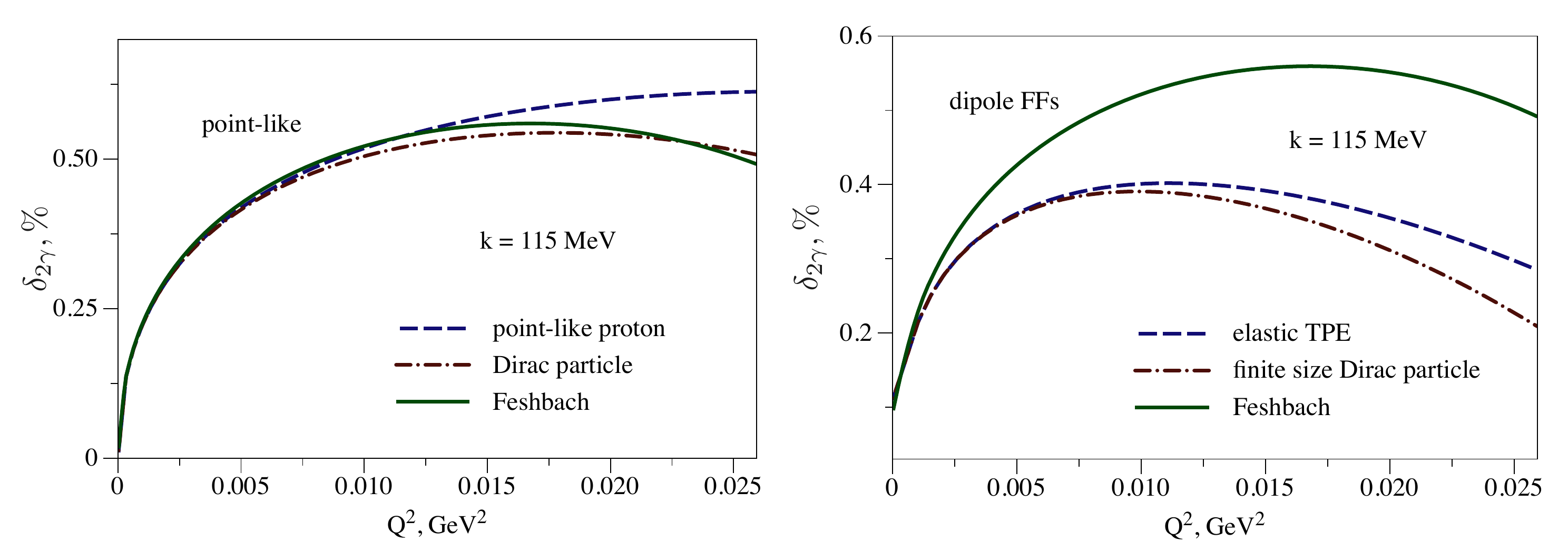}}
\caption{Left panel: TPE correction for the case of a point-like proton, compared with the case when one neglects the magnetic moment (Dirac particle), as well as the Feshbach result (corresponding with Coulomb photon exchange). Right panel: TPE correction for the case of the proton with electric and magnetic form factors of the dipole form. We compare the box graph calculation with the Feshbach term corrected by the recoil correction $ 1 + m / M$.} 
\label{MUSE_region_Feshbach_term}
\end{figure}

We show in Fig. \ref{MUSE_region_Feshbach_term} (left panel) the comparison between the Feshbach term, the TPE contribution for point-like Dirac particles, and the TPE for a point-like proton, with inclusion of the magnetic moment contribution. It is seen that the Feshbach correction of Eq. (\ref{full_feshbach}) with account of the recoil correction factor $ \left( 1 + m/M \right) $ describes the result for point-like Dirac particles quite well in the kinematics of the MUSE experiment.

We also show in Fig. \ref{MUSE_region_Feshbach_term} (right panel) the effect of the proton FFs, according to the full numerical calculation of Ref. \cite{Tomalak:2014dja}. In the low $Q^2$ kinematics of the MUSE experiment, the inclusion of the FFs provides a reduction of the TPE by around $ 40~\% $ at $ Q^2 \approx 0.025 ~\mathrm{GeV}^2 $, consequently one should use the full numerical calculation of Ref. \cite{Tomalak:2014dja} (corresponding with the elastic TPE result in Fig. \ref{MUSE_region_Feshbach_term}) in MUSE kinematics.

\section{Forward unpolarized doubly virtual Compton scattering tensor}
\label{sec3}

The TPE contribution, $ \delta_{2 \gamma}$, for the muon-proton scattering process is in general given by the interference of the one-photon exchange amplitude ($ T^{\mathrm{OPE}} $) and the two-photon exchange amplitude ($ T^{\mathrm{TPE}}$) as
\ber \label{TPE_correction}
\delta_{2 \gamma} = \frac{2 \Re \left(\sum \limits_{\mathrm{spin}}T^{\mathrm{TPE}} \left( T^{\mathrm{OPE}} \right)^* \right)}{\sum \limits_{\mathrm{spin}}  |T^{\mathrm{OPE}}|^2}. 
\eer
The OPE expression in the denominator of Eq.~(\ref{TPE_correction}) is given by~:
\ber \label{OPE}
\sum \limits_{\mathrm{spin}}  |T^{\mathrm{OPE}}|^2 = \frac{8 e^4}{\tau} \frac{1 - \varepsilon_0}{1 - \varepsilon} 
\left( \varepsilon G^2_E(Q^2) + \tau G^2_M (Q^2) \right),  
\eer
where $G_E$ ($G_M$) are the proton electric (magnetic) form factors respectively, and with kinematical quantities $\tau$ and $\varepsilon_0$ defined as in Eq.~(\ref{eq:deftau}). The interference between the OPE and TPE amplitudes in the numerator of Eq.~(\ref{TPE_correction}) can be expressed as
\ber
\label{OPETPE}
&&2 \Re \left(\sum \limits_{\mathrm{spin}}T^{\mathrm{TPE}} \left( T^{\mathrm{OPE}} \right)^* \right)  =  -  \Re  \frac{4 \pi  e^4}{Q^2}  \mathop{\mathlarger{\int}} \frac{ i \mathrm{d}^4 \tilde{q}}{\left( 2 \pi \right)^4} \frac{L^{\mu \nu \alpha} H_{\mu \nu \alpha}  }{\left(\tilde{q} - \frac{q}{2} \right)^2 \left(\tilde{q} + \frac{q}{2} \right)^2 } , \nonumber \\
&&L^{\mu \nu \alpha}  =  \textbf{Tr}  \left\{  \left( \gamma^\mu \frac{\hat{K}-\hat{\tilde{q}}+m}{\left(K-\tilde{q}\right)^2-m^2} \gamma^\nu + \gamma^\nu \frac{\hat{K}+\hat{\tilde{q}}+m}{\left(K+\tilde{q}\right)^2-m^2} \gamma^\mu  \right) (\hat{k}+m ) \gamma^\alpha (\hat{k^\prime}+m ) \right\}, \nonumber \\
&&H^{\mu \nu \alpha}  =  \textbf{Tr} \left\{ M^{\mu \nu} \left(\hat{p}+M \right) \Gamma^\alpha ( Q^2) (\hat{p^\prime}+M )   \right\}, 
\eer
with all momenta defined as in Fig.~\ref{TPE_kinematics}, where $ \tilde{q} $ is the loop-momentum over which one integrates, and where $\Gamma^\alpha$ denotes the on-shell proton electromagnetic vertex:
\beq
\label{emvertex}
\Gamma^\alpha ( Q^2 ) \;=\; F_D (Q^2) \, \gamma^\alpha \;+\;
F_P(Q^2) \, \frac{i \sigma^{\alpha \beta} q_\beta}{2 M} \, ,
\eeq
with $F_D$ ($F_P$) the Dirac (Pauli) form factors of the proton respectively.
Furthermore in Eq.~(\ref{OPETPE}), $M^{\mu \nu}$ denotes the proton doubly virtual Compton scattering tensor. 

The main aim of the present work is to quantitatively estimate the {\it inelastic} TPE contribution in the low $ Q^2 $ region, corresponding with the MUSE kinematics. 
For this purpose we will approximate the hadronic tensor $M^{\mu \nu}$ in Eq.~(\ref{OPETPE}) 
by the forward doubly virtual scattering (VVCS) tensor. The unpolarized forward VVCS process $ \gamma^\ast (\tilde{q}) + p (P) \to \gamma^\ast (\tilde{q}) + p (P) $ is described by two invariant amplitudes $ \mathrm{T}_1 $ and $ \mathrm{T}_2 $, which are defined in this work through the tensor decomposition:
\ber
\label{vvcs_tensor}
M^{\mu \nu}  = 
 - \left( -g^{\mu\nu}+\frac{\tilde{q}^{\mu} \tilde{q}^{\nu}}{\tilde{q}^2 }\right)
\mathrm{T}_1 (\tilde{\nu}, \tilde{Q}^2 ) 
 - \frac{1}{M^2} \left(P^{\mu}-\frac{ M \tilde{\nu} }{\tilde{q}^2 }\,\tilde{q}^{\mu}\right) \left(P^{\nu}-\frac{ M \tilde{\nu} }{\tilde{q}^2 }\, \tilde{q}^{\nu} \right) \mathrm{T}_2 (\tilde{\nu}, \tilde{Q}^2 ),
\eer%
with the photon energy $ \tilde{\nu} = \left( P \cdot \tilde{q} \right)/M $ and the squared photon virtuality $ \tilde{Q}^2 \equiv - \tilde{q}^2 $. The absorptive parts of the amplitudes $ \mathrm{T}_1 $ and $ \mathrm{T}_2 $ are related to the proton structure functions $ F_1 $ and $ F_2 $ by
\ber \label{imaginary_ampl}
\Im \mathrm{T}_1(\tilde \nu, \tilde Q^2)  =  \frac{e^2}{4 M} F_1(\tilde \nu, \tilde Q^2), \qquad
\Im \mathrm{T}_2 (\tilde \nu, \tilde Q^2)  =  \frac{e^2}{4 \tilde{\nu}} F_2(\tilde \nu, \tilde Q^2).
\eer

In this work, we will approximate the unpolarized tensor $M^{\mu \nu}$ entering Eq.~(\ref{OPETPE})  for the process 
$\gamma^\ast (q_1=\tilde q + q/2) + p (P - q/2) \to \gamma^\ast (q_2 = \tilde q - q/2) + p (P + q/2)$  
in the low momentum transfer limit $q \to 0$, i.e. $q_1 \approx q_2$,  
by~\cite{Tomalak:2015aoa}:
\ber
\label{vvcs_agreement_with_elastic1}
M^{\mu \nu}& \approx &
- \left( -g^{\mu\nu}+\frac{q_1^{\mu} q_2^{\nu}}{q_1 \cdot q_2 }\right)
T_1\left(\tilde{\nu}, - q_1 \cdot q_2 \right) \nonumber \\
& & - \frac{1}{M^2} \left(P^{\mu}-\frac{ M \tilde{\nu} }{q_1 \cdot q_2 }\, q_1^{\mu}\right) \left(P^{\nu}-\frac{ M \tilde{\nu} }{q_1 \cdot q_2 }\, q_2^{\nu} \right) T_2 \left(\tilde{\nu}, - q_1 \cdot q_2\right) . 
\eer
By using the electromagnetic gauge invariance of the lepton tensor,
\ber
q_1^\nu L_{\mu \nu \alpha} = 0, \quad q_2^\mu L_{\mu \nu \alpha} = 0, 
\eer
the hadronic tensor of Eq.~(\ref{vvcs_agreement_with_elastic1}) can be expressed equivalently as
\ber
\label{vvcs_agreement_with_elastic2}
M^{\mu \nu}& \approx &
- \left( -g^{\mu\nu}+\frac{q^{\mu} q^{\nu}}{\tilde{Q}^2 - \frac{Q^2}{4} }\right)
T_1\left(\tilde{\nu}, \tilde{Q}^2 - \frac{Q^2}{4}\right) \nonumber \\
& & - \frac{1}{M^2} \left(P^{\mu}+\frac{ M \tilde{\nu} }{\tilde{Q}^2 - \frac{Q^2}{4} }\,q^{\mu}\right) \left(P^{\nu}-\frac{ M \tilde{\nu} }{\tilde{Q}^2 - \frac{Q^2}{4} }\, q^{\nu} \right) T_2 \left(\tilde{\nu}, \tilde{Q}^2 - \frac{Q^2}{4}\right) , 
\eer
which is the tensor form which we will use in our evaluations of 
$\delta_{2 \gamma}$.

The real part of the amplitude $ \mathrm{T}_1 $ can be expressed through a subtracted dispersion relation (DR) as integral over the invariant mass $ W^2$ of the intermediate hadronic state  as
\ber \label{forward_VVCS_DR_T1}
\Re \mathrm{T}_1 (\tilde{\nu}, \tilde{Q}^2) &  = &  \Re \mathrm{T}^{\mathrm{Born}}_1 ( \tilde{\nu}, \tilde{Q}^2 ) + \mathrm{T}^{\mathrm{subt}}_1 (0, \tilde{Q}^2) \nonumber \\
& + & \frac{2 \tilde{\nu}^2 }{\pi }  \mathop{\mathlarger{\int}} \limits^{~~ \infty}_{W^2_{\mathrm{thr}}} \frac{ e^2 M F_1\left((W^2 - P^2 + \tilde Q^2)/(2 M),\tilde{Q}^2\right) \mathrm{d} W^2}{\left( W^2 - P^2 + \tilde{Q}^2 \right)  \left( \left( P + \tilde{q} \right)^2 - W^2 + i \varepsilon \right) \left( \left( P - \tilde{q} \right)^2 - W^2 + i \varepsilon \right)} ,
\eer
with the pion-proton inelastic threshold: $ W^2_{\mathrm{thr}} = \left( M + m_{\pi} \right)^2  \approx 1.15 ~\mathrm{GeV}^2 $, where $ m_{\pi} $ denotes the pion mass, and where $  \mathrm{T}^{\mathrm{subt}}_1 (0, \tilde{Q}^2 ) $ is the subtraction function at $ \tilde{\nu} = 0$. The real part of the amplitude $  \mathrm{T}_2 $ can be obtained from an unsubtracted DR:
\ber \label{forward_VVCS_DR_T2}
\Re \mathrm{T}_2 (\tilde{\nu}, \tilde{Q}^2) &  = & \Re \mathrm{T}^{\mathrm{Born}}_2 ( \tilde{\nu}, \tilde{Q}^2 ) +  \frac{1 }{\pi}   \mathop{\mathlarger{\int}} \limits^{~~ \infty}_{W^2_{\mathrm{thr}}} \frac{ e^2 M F_2 \left((W^2 - P^2 + \tilde Q^2)/(2M), \tilde{Q}^2\right)  \mathrm{d} W^2}{\left( \left( P + \tilde{q} \right)^2 - W^2 + i \varepsilon \right) \left( \left( P - \tilde{q} \right)^2 - W^2 + i \varepsilon \right)},
\eer
In Eqs. (\ref{forward_VVCS_DR_T1}) and (\ref{forward_VVCS_DR_T2}), the Born contributions to the unpolarized Compton amplitudes $ \mathrm{T}^{\mathrm{Born}}_1 $ and $ \mathrm{T}^{\mathrm{Born}}_2  $, due to the proton intermediate state, are given by
\ber
\Re \mathrm{T}_1^{\mathrm{Born}} ( \tilde{\nu}, \tilde{Q}^2 )  & = & \frac{\alpha}{M} \left(  \frac{\tilde{Q}^4 G^2_M(\tilde{Q}^2)}{  \tilde{Q}^4  - 4 M^2 \tilde{\nu}^2 } -F_D^2 (\tilde{Q}^2) \right), \label{t1_Born} \\
\Re \mathrm{T}_2^{\mathrm{Born}} ( \tilde{\nu}, \tilde{Q}^2 )  & = &  4 M \alpha \tilde{Q}^2 \frac{ F_D^2(\tilde{Q}^2) + \frac{\tilde{Q}^2}{4 M^2} F^2_P (\tilde{Q}^2) }{ \tilde{Q}^4  - 4 M^2 \tilde{\nu}^2}. \label{t2_Born}
\eer
Note that in the derivation of a DR as given e.g. in Eq. (\ref{forward_VVCS_DR_T1}) the elastic (nucleon pole) term contribution, given by only the first term of Eq. (\ref{t1_Born}), correctly appears. This pole contribution differs from the Born term by
\ber \label{nonpole}
\mathrm{T}_1^{\mathrm{Born}} ( \tilde{\nu}, \tilde{Q}^2 ) - \mathrm{T}_1^{\mathrm{pole}} ( \tilde{\nu}, \tilde{Q}^2 )  = - \frac{\alpha}{M} F^2_D ( \tilde{Q}^2 ).
\eer
As this is an energy ($ \tilde{\nu} $) independent function, we have absorbed it in the definition of $ \mathrm{T}^{\mathrm{subt}}_1 ( 0, \tilde{Q}^2 ) $, which in Eq. (\ref{forward_VVCS_DR_T1}) is defined as
\ber \label{floop}
\mathrm{T}^{\mathrm{subt}}_1 (0, \tilde{Q}^2 ) \equiv \mathrm{T}_1 (0, \tilde{Q}^2) - \mathrm{T}^{\mathrm{Born}}_1 (0, \tilde{Q}^2)  \equiv  \tilde{Q}^2 \beta (\tilde{Q}^2).
\eer
The advantage of expressing the amplitude $ \mathrm{T}_1 $ w.r.t. to its Born contribution, results from the fact that the non-Born amplitude in Eq. (\ref{floop}) starts at $\tilde{Q}^2$, and is usually parametrized in terms of polarizabilities, i. e. the function $ \beta (\tilde{Q}^2) $ at $ \tilde{Q}^2 = 0 $ is given by the magnetic polarizability $ \beta_M $: $ \beta ( 0 ) = \beta_M $ \cite{Birse:2012eb,Carlson:2011zd}.

In order to evaluate the inelastic TPE contribution using the forward non-Born VVCS amplitudes, we will need the information on the proton structure functions $ F_1 $ and $ F_2 $ as well as to specify the subtraction function in Eq. (\ref{floop}) which is parametrized through the function $ \beta ( \tilde{Q}^2 ) $. We will evaluate the subtraction function contribution in Section \ref{sec4}, and the dispersive contribution due to the structure functions $ F_1 $ and $ F_2 $ in Sec. \ref{sec5}.

\section{Subtraction function contribution to TPE correction}
\label{sec4}

In the present section we will discuss the TPE correction due to the subtraction function $\mathrm{T}^{\mathrm{subt}}_1 (0, \tilde{Q}^2 )$. For this purpose, we will compare three different estimates for $ \beta ( \tilde{Q}^2 ) $ defined through Eq.~(\ref{floop}). At low $ Q^2 $ we will use existing estimates from heavy-baryon and baryon chiral perturbation theory. Furthermore, we will provide an empirical determination of $ \beta ( \tilde{Q}^2 ) $ based on the high-energy behavior of the Compton amplitude.

\subsection{Heavy-baryon ChPT subtraction function}
\label{sec4a}

\begin{figure}[h]
\centering{\includegraphics[width=0.75\textwidth]{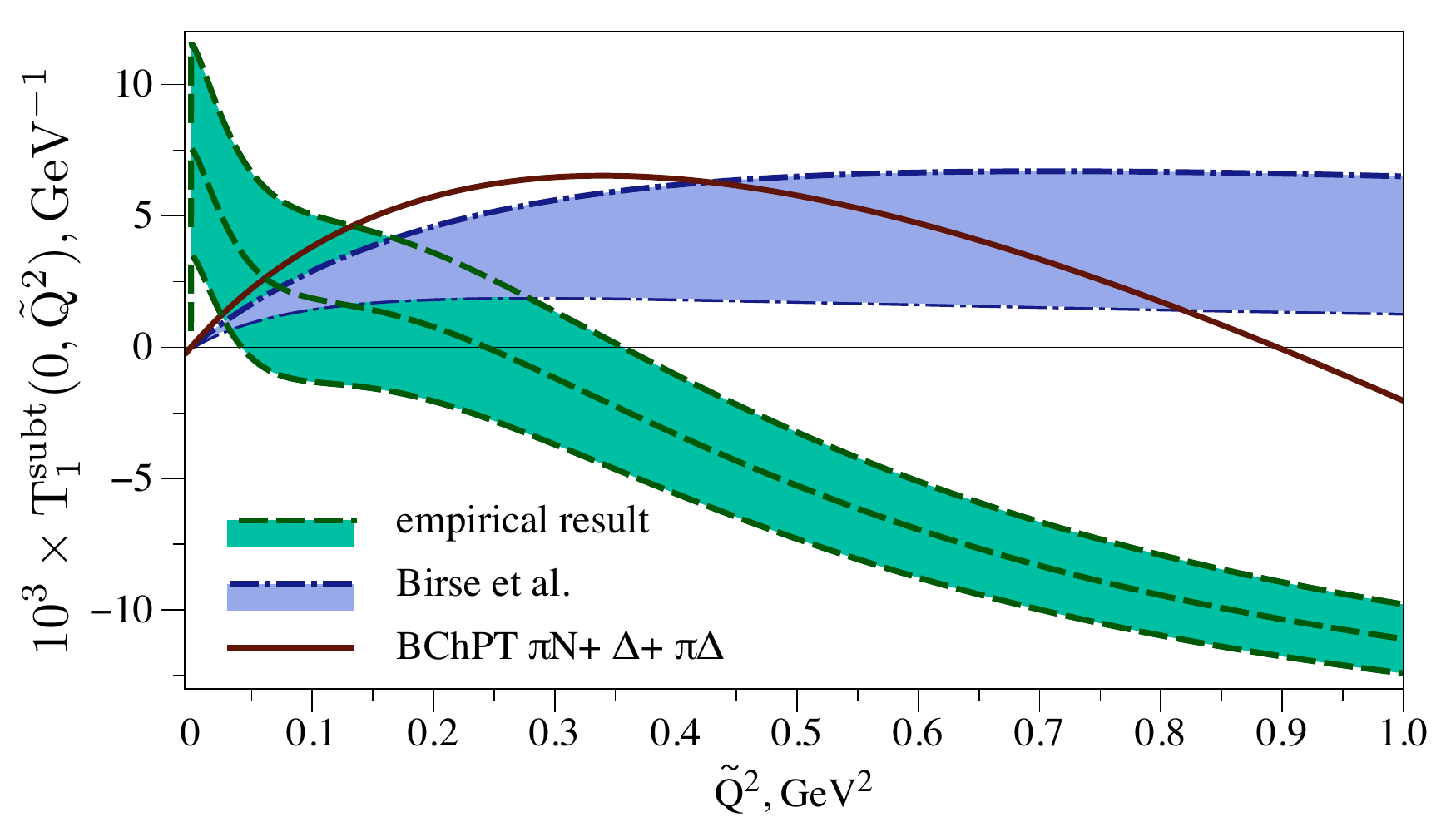}}
\caption{The empirical subtraction function of Eq. (\ref{subtraction_from_HE}) in comparison with the subtraction functions from HBChPT of Birse et al.~\cite{Birse:2012eb}, and from BChPT~\cite{Alarcon:2013cba}.} 
\label{experimental_T1}
\end{figure}

\begin{figure}[h]
\centering{\includegraphics[width=0.75\textwidth]{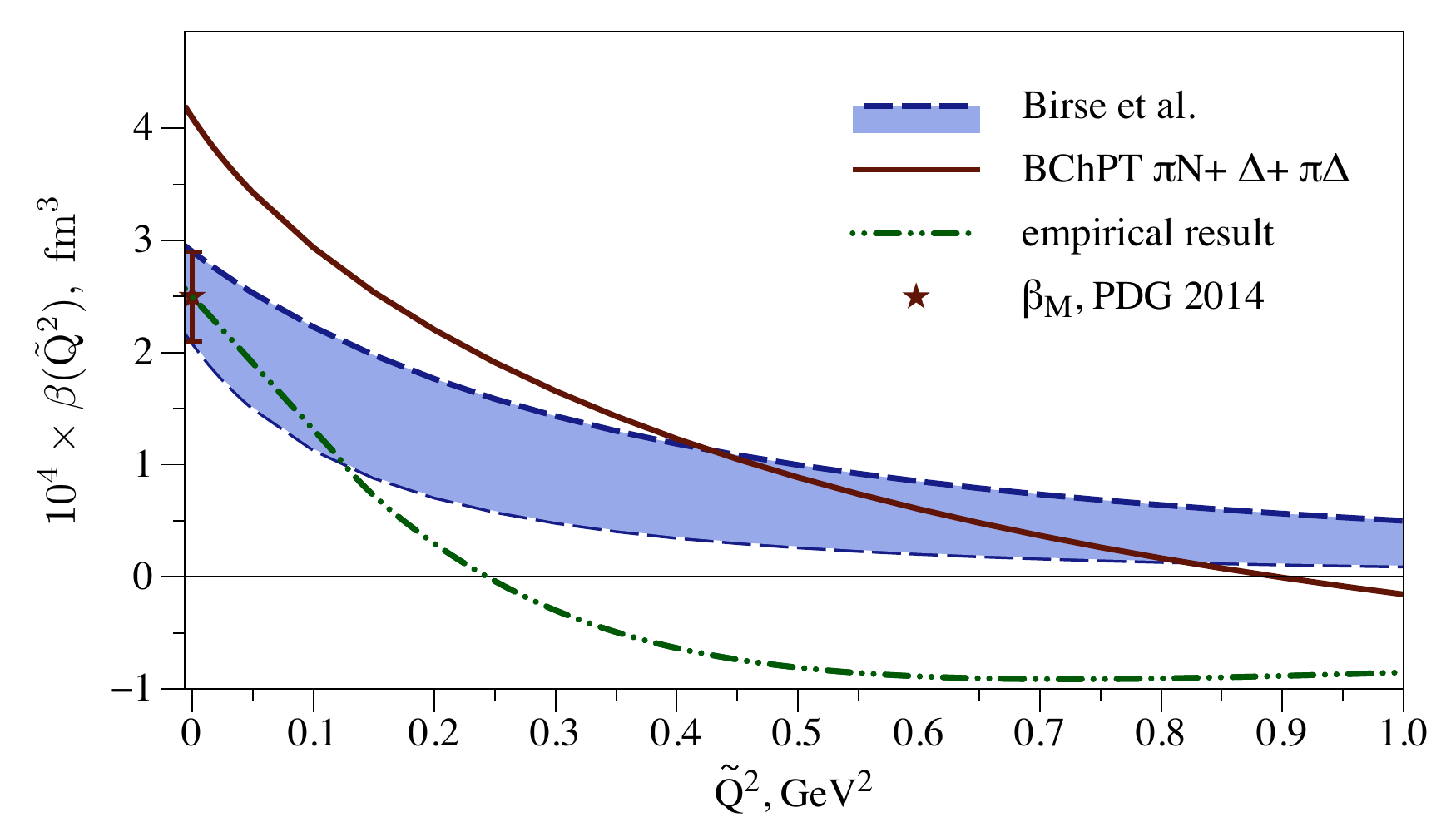}}
\caption{The empirical estimate for the magnetic polarizability $ \beta(\tilde{Q}^2) $ based on Eqs. (\ref{floop}, \ref{subtraction_from_HE}) compared with the HBChPT result of Birse et al.~\cite{Birse:2012eb} normalized to the PDG value $ \beta(0) =  \left( 2.5 \pm 0.4 \right)  \times 10^{-4} ~\mathrm{fm^3}$~\cite{Agashe:2014kda}, and with the BChPT result~\cite{Alarcon:2013cba}.} 
\label{beta_Q2}
\end{figure}

First of all, we show the fit of Ref.~\cite{Birse:2012eb} obtained by matching the heavy-baryon chiral perturbation theory (HBChPT) result to a dipole behavior:
\ber \label{Birse_polarizability}
\beta (\tilde{Q}^2 ) = \frac{\beta_M}{\left( 1 + \tilde{Q}^2 / \Lambda^2 \right)^2}, \qquad \Lambda =  530 - 842  ~\mathrm{MeV},
\eer
with the value of the magnetic polarizability $ \beta_M = (2.5 \pm 0.4) \times 10^{-4} ~\mathrm{fm^3}$ taken from PDG \cite{Agashe:2014kda}. For the purpose of showing error bands in our numerical estimates, we choose the lower and upper edges of such bands to correspond with the values: $ \Lambda = 530 ~\mathrm{MeV},~\beta_M = 2.1 \times 10^{-4} ~\mathrm{fm^3} $ and $ \Lambda = 842 ~\mathrm{MeV},~\beta_M = 2.9 \times 10^{-4} ~\mathrm{fm^3} $ respectively. The resulting bands for $ \mathrm{T}^{\mathrm{subt}}_1 (0, \tilde{Q}^2 ) $ are shown in Fig. \ref{experimental_T1}, and correspondingly for $ \beta(\tilde{Q}^2) $ in Fig. \ref{beta_Q2} (blue bands). 

\subsection{Baryon ChPT subtraction function}
\label{sec4b}

Second, we also show the prediction for $ \beta ( \tilde{Q}^2 ) $ resulting from the covariant baryon chiral perturbation theory (BChPT)~\cite{Alarcon:2013cba}, with $\beta$ decomposed as
\ber \label{Alarcon_polarizability}
\beta (\tilde{Q}^2 ) = \beta_{\pi N} (\tilde{Q}^2 ) + \beta_{\Delta} (\tilde{Q}^2 ) + \beta_{\pi \Delta} (\tilde{Q}^2 ), 
\eer
with $ \beta_{\pi N} (\tilde{Q}^2 ) $ the $ \mathcal{O} ( p^3 ) $ diamagnetic polarizability contribution from $ \pi N $ loops given by Eq. (22) of Ref. \cite{Alarcon:2013cba},
$ \beta_{\Delta} (\tilde{Q}^2 )  $ the paramagnetic contribution of the $ \Delta $-resonance to the magnetic polarizability \cite{Lensky:2015awa} and  $ \beta_{\pi \Delta} (\tilde{Q}^2 ) $ the $ \mathcal{O} ( p^{7/2} ) $ at $ p \simeq m_\pi $ diamagnetic polarizability contribution from $ \pi \Delta $ loops \cite{Lensky:2015awa} . 

In Fig. \ref{experimental_T1}, we compare the heavy-baryon and baryon ChPT predictions for $ \mathrm{T}^{\mathrm{subt}}_1 (0, \tilde{Q}^2 ) $. Notice that the HBChPT value of $ \beta(0) $ is taken from a fit to data (PDG 2014) whereas the baryon ChPT value of $ \beta (0) $ results from the sum of the positive paramagnetic part due to the $s$-channel $ \Delta$-excitation $ \beta_{\Delta} (0) \simeq 7 \times 10^{-4}~\mathrm{fm}^3 $, and the negative diamagnetic part due to $ \pi N $ and $ \pi \Delta $ loops, i.e. $ \beta_{\pi N} (0) = - 2 \times 10^{-4}~\mathrm{fm}^3 $ and $ \beta_{\pi \Delta} (0) = - 1.2 \times 10^{-4}~\mathrm{fm}^3 $.

\subsection{Empirical determination of the subtraction function}
\label{sec4c}

In this section, we discuss an empirical estimate of the function $ \beta ( \tilde{Q}^2 ) $ at non-zero $ \tilde{Q}^2 $ from experimental information on inelastic electron-proton scattering. Following the idea of Refs. \cite{Gasser:1974wd,Gasser:2015dwa}, the subtraction function can be obtained from an unsubtracted dispersion relation for the amplitude $ \mathrm{T}_1 ( \tilde{\nu}, \tilde{Q}^2 ) - \mathrm{T}_1^\mathrm{R} ( \tilde{\nu}, \tilde{Q}^2 ) $, where $\mathrm{T}_1^\mathrm{R}$ denotes a Regge amplitude which is chosen such as to match the high-energy behavior of the amplitude $ \mathrm{T}_1$, i.e. 
$\mathrm{T}_1 - \mathrm{T}_1^\mathrm{R} \rightarrow 0$ for $\tilde \nu \rightarrow \infty$  \footnote{This assumes that $ T_1 (\tilde{\nu}, \tilde{Q}^2) $ does not have a fixed-pole behavior when $\tilde \nu \rightarrow \infty$ . }. 
The function $\mathrm{T}_1^\mathrm{R}$ is chosen as a sum over the leading Regge trajectories:
\ber \label{regge_T1}
\mathrm{T}_1^{\mathrm{R}} ( \tilde{\nu}, \tilde{Q}^2 ) & \equiv & -  \frac{\pi \alpha}{M}  \sum_{\alpha_0 > 0} \frac{\gamma_{\alpha_0} (\tilde{Q}^2 )}{\sin \pi \alpha_0}  \left \{ \left(\tilde{\nu}_0 - \tilde{\nu} - i \varepsilon \right)^{\alpha_0} + \left( \tilde{\nu}_0 + \tilde{\nu} - i \varepsilon \right)^{\alpha_0}  \right \} \nonumber \\
& - &  \frac{\pi \alpha}{M}  \sum_{{\alpha_0}  > 1} \frac{  {\alpha_0}  \tilde{\nu}_0  \gamma_{\alpha_0}  (\tilde{Q}^2 )}{\sin \pi \left({\alpha_0}  - 1 \right)}  \left \{ \left(\tilde{\nu}_0 - \tilde{\nu} - i \varepsilon \right)^{{\alpha_0} - 1} + \left(\tilde{\nu}_0 + \tilde{\nu} - i \varepsilon \right)^{{\alpha_0}  - 1} \right \},
\eer
with the intercept $ {\alpha_0} > 0 $, $ \tilde{\nu}_0 $ is a reference hadronic scale which is used as a free parameter and $  \gamma_{\alpha_0} (\tilde{Q}^2 ) $ are the Regge residues. Using Eq. (\ref{imaginary_ampl}), the imaginary part of $ \mathrm{T}^\mathrm{R}_1 $ yields the corresponding Regge structure:
\ber \label{F1R}
F_1^{\mathrm{R}} ( \tilde{\nu}, \tilde{Q}^2 ) \equiv \frac{M}{\pi \alpha} \Im \mathrm{T}^{R}_1 ( \tilde{\nu}, \tilde{Q}^2 ) & = & \sum_{{\alpha_0} > 0} \gamma_{\alpha_0} (\tilde{Q}^2 )  \left( \tilde{\nu} - \tilde{\nu}_0 \right)^{{\alpha_0}} \Theta \left( \tilde{\nu} - \tilde{\nu}_0  \right) \nonumber \\
& + & \sum_{{\alpha_0} > 1}  \gamma_{\alpha_0} (\tilde{Q}^2 ) {\alpha_0} \tilde{\nu}_0  \left( \tilde{\nu} - \tilde{\nu}_0 \right)^{{\alpha_0}-1} \Theta \left( \tilde{\nu} - \tilde{\nu}_0  \right).
\eer
The Regge residues $ \gamma_{\alpha_0} (\tilde{Q}^2 ) $ can be obtained by performing a fit to inclusive electroproduction data on a proton. In our work we use the Donnachie-Landshoff (DL) high-energy fit \cite{Donnachie:2004pi} to obtain the proton structure function $F_1$ as
\ber \label{F1HE}
F_1 ( \tilde{\nu}, \tilde{Q}^2 )  \underset{ \tilde{\nu}\gg}{ \longrightarrow}  \sum \limits_{\alpha_0 > 0} \gamma_{\alpha_0} ( \tilde{Q}^2 )   \tilde{\nu} ^{\alpha_0},
\eer
where the values of the Regge intercepts $ \alpha_0 $ and the residue functions $ \gamma_{\alpha_0} (\tilde{Q}^2 ) $ are detailed in Appendix \ref{experimental_T1_details}.

By comparing Eq. (\ref{F1R}) and (\ref{F1HE}) we notice that the second term in Eq. (\ref{F1R}) is chosen such that for the Regge trajectory with $ 1 < \alpha_0 < 2 $ ("Pomeron"):
\ber
F_1 (\tilde{\nu}, \tilde{Q}^2 ) - F^R_1 (\tilde{\nu}, \tilde{Q}^2 )  \underset{\tilde{\nu}\gg}{ \sim}  \tilde{\nu} ^{\alpha_0-2},
\eer
whereas for the Regge trajectory with $ 0 < \alpha_0 < 1 $ (Reggeon):
\ber
F_1 (\tilde{\nu}, \tilde{Q}^2 ) - F^R_1 (\tilde{\nu}, \tilde{Q}^2 ) \underset{\tilde{\nu}\gg}{ \sim}  \tilde{\nu} ^{\alpha_0-1}.
\eer
This ensures that in all cases the quantity $[ F_1 ( \tilde{\nu}, \tilde{Q}^2 ) - F_1^{\mathrm{R}} ( \tilde{\nu}, \tilde{Q}^2 ) ] \rightarrow 0$ 
when $\tilde \nu \rightarrow \infty$. 

Consequently, one can write down an unsubtracted dispersion relation for $\mathrm{T}_1 - \mathrm{T}_1^\mathrm{R}$ at fixed $ \tilde{Q}^2$ as
\ber \label{subtraction_from_HE0}
\mathrm{T}_1 ( \tilde{\nu}, \tilde{Q}^2 ) - \mathrm{T}_1^\mathrm{R}  ( \tilde{\nu}, \tilde{Q}^2 ) = \mathrm{T}_1^\mathrm{pole} ( \tilde{\nu}, \tilde{Q}^2 ) + \frac{2}{\pi}  \mathop{\mathlarger{\int}}  \mathrm{d} \nu' \frac{\nu' \Im \left [ \mathrm{T}_1  ( \nu', \tilde{Q}^2 ) - \mathrm{T}_1^\mathrm{R}  ( \nu', \tilde{Q}^2 ) \right ]}{\nu'^2 - \tilde{\nu}^2} .
\eer
Using Eqs.~(\ref{imaginary_ampl}, \ref{nonpole}, \ref{floop}), 
this yields an expression for $ \mathrm{T}^{\mathrm{subt}}_1 ( 0, \tilde{Q}^2 ) $ which expressed in terms of $ W^2 \equiv 2 M \nu^\prime + M^2 - \tilde{Q}^2 $ is given by
\ber \label{subtraction_from_HE}
&&\mathrm{T}^{\mathrm{subt}}_1 ( 0, \tilde{Q}^2 ) = \mathrm{T}_1^\mathrm{R}  ( 0, \tilde{Q}^2 ) + \frac{\alpha}{M} F^2_D ( \tilde{Q}^2 ) 
\nonumber \\
&& + \frac{2 \alpha}{ M}  \mathop{\mathlarger{\int}} \limits^{~~ \infty}_{s_{\mathrm{thr}}} 
\frac{ F_1\left((W^2 - M^2 + \tilde Q^2)/(2 M),\tilde{Q}^2 \right) - F^R_1\left((W^2 - M^2 + \tilde Q^2)/(2 M),\tilde{Q}^2 \right) }{ W^2 - M^2 + \tilde{Q}^2} \mathrm{d} W^2, \quad \quad
\eer
where the lower integration limit in Eq. (\ref{subtraction_from_HE}) $s_{\mathrm{thr}}$ is given by
\ber
 s_{\mathrm{thr}} = \mathrm{\mathrm{min}} \left( s_0  \equiv 2 M \tilde{\nu}_0 + M^2 - \tilde{Q}^2,~ W^2_{\mathrm{thr}} = ( M + m_\pi)^2 \right),
 \eer
 corresponding with a branch cut of $F_1$ starting at $W_{\mathrm{thr}}^2$ and a branch cut of $F_1^{\mathrm{R}}$ starting at $s_0$. 
 Eq. (\ref{subtraction_from_HE}) allows to quantitatively estimate the subtraction function given the structure function $ F_1 $, the Regge fit determining $ F^R_1 $ of the form of Eq. (\ref{F1R}), as well as the corresponding value of $\mathrm{T}^\mathrm{R}_1 ( 0, \tilde{Q}^2 ) $ which follows from Eq. (\ref{regge_T1}) as
\ber \label{T10Q2}
\mathrm{T}^\mathrm{R}_1 (0, \tilde{Q}^2 ) & = & - \frac{2 \pi \alpha}{M} \sum \limits_{\alpha_0 > 0} \frac{\gamma_{\alpha_0} ( \tilde{Q}^2 )}{\sin \pi \alpha_0}  \tilde{\nu}_0^{\alpha_0} - \frac{2 \pi \alpha}{M} \sum \limits_{\alpha_0 > 1} \frac{ \alpha_0 \tilde{\nu}_0 \gamma_{\alpha_0} ( \tilde{Q}^2 )}{\sin \pi \left( \alpha_0 - 1 \right) }  \tilde{\nu}_0^{\alpha_0 - 1 },
\eer
and is also fully determined by the Regge fit. 

\begin{figure}[h]
\centering{\includegraphics[width=1\textwidth]{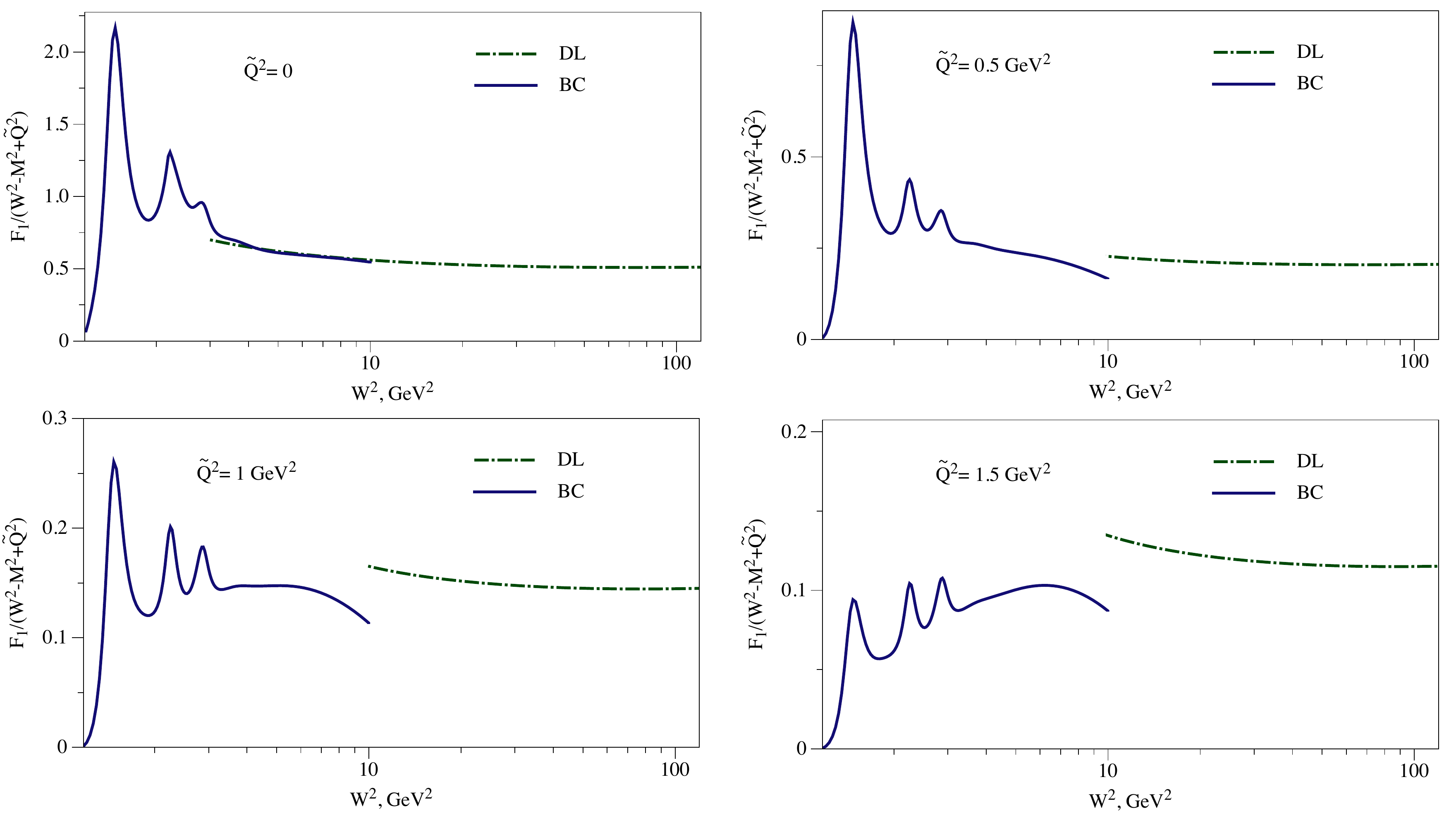}}
\caption{Fits of the proton structure function $ F_1 $ used in our estimates entering the dispersion integral in  Eq.~(\ref{subtraction_from_HE}).}
\label{F1_plots}
\end{figure}

In our numerical evaluation of Eq.~(\ref{subtraction_from_HE}), we describe the proton structure function $ F_1 $ in the resonance region by the fit performed by Christy and Bosted (BC) \cite{Christy:2007ve}. This fit is valid in the following region of kinematic variables: $ 0 < \tilde{Q}^2 < 8 ~\mathrm{GeV}^2 $, and $W^2 < 9.61 ~\mathrm{GeV}^2 \approx 10 ~\mathrm{GeV}^2 $. For the dispersion integral in Eq. (\ref{subtraction_from_HE}) we connect the BC fit with the DL high-energy fit starting from $ W^2 = 10 ~\mathrm{GeV}^2 $. 
The latter fit is described in Appendix \ref{experimental_T1_details}. 
The resulting proton structure function $ F_1 $ is shown in Fig. \ref{F1_plots} as it enters the integral of Eq.~(\ref{subtraction_from_HE}). We add a $ 3~\% $ error band to the BC fit \cite{Christy:2007ve} and use the same error estimate for all Regge pole residues. We notice that at low values of $\tilde Q^2$, both fits either overlap or are very close 
around the matching point $W^2 \approx 10$ GeV$^2$. With increasing values of $\tilde{Q}^2$ there is a slight mismatch in both fits around $W^2 = 10$ GeV$^2$, which is due to the fact that the BC fit has not accounted for the HERA high-energy data, and the DL fit has not accounted for the lower $W$ data. Even though a combined fit of all data would be very worthwhile, or a smooth interpolating procedure between the BC and DL fits could easily be performed, for our purpose we will only need data at lower value of $\tilde Q^2$ up to about 1 GeV$^2$. For this purpose, we can just split the $W^2$ integral entering Eq.~(\ref{subtraction_from_HE}) in a region $W^2 < 10$ GeV$^2$ where we will use the BC fit and a region $W^2 > 10$ GeV$^2$ where we will use the DL fit. 

\begin{figure}[h]
\centering{\includegraphics[width=1.\textwidth]{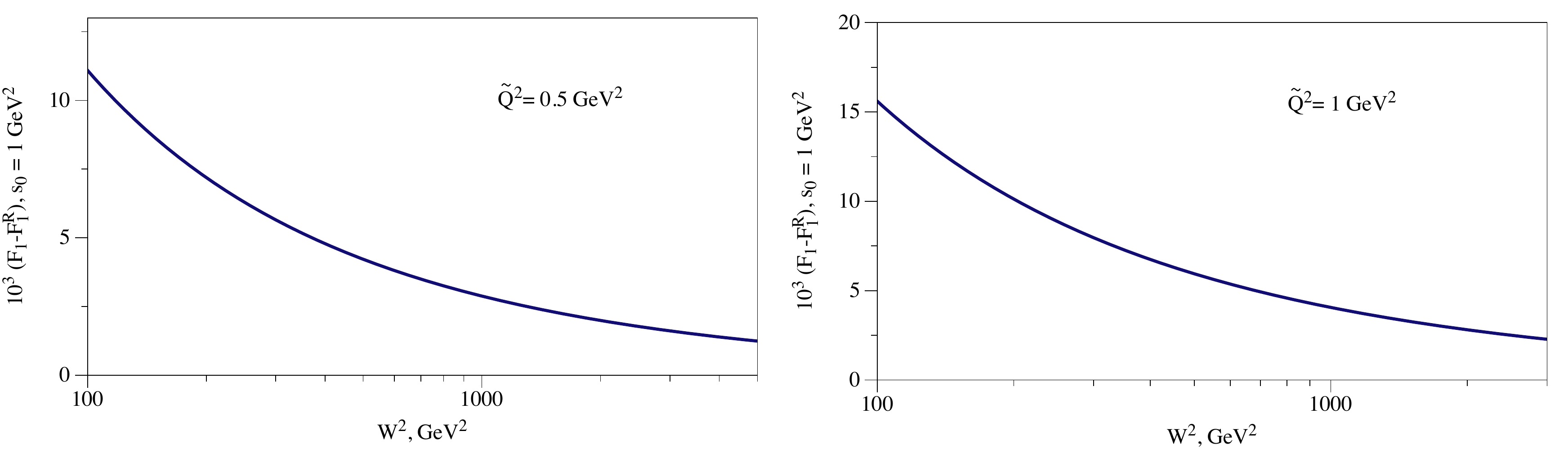}}
\caption{High-energy behavior of the function $ F_1 - F_1^{\mathrm{R}} $ for the fixed value $ s_0 =  1 ~\mathrm{GeV}^2$.}
\label{F1_HE}
\end{figure}

In Fig. \ref{F1_HE}, we demonstrate explicitly the vanishing high-energy behavior of the quantity $F_1 - F_1^{\mathrm{R}}$, which 
is the necessary condition for the unsubtracted DR of Eq. (\ref{subtraction_from_HE}) to hold.

\begin{figure}[h]
\centering{\includegraphics[width=.7\textwidth]{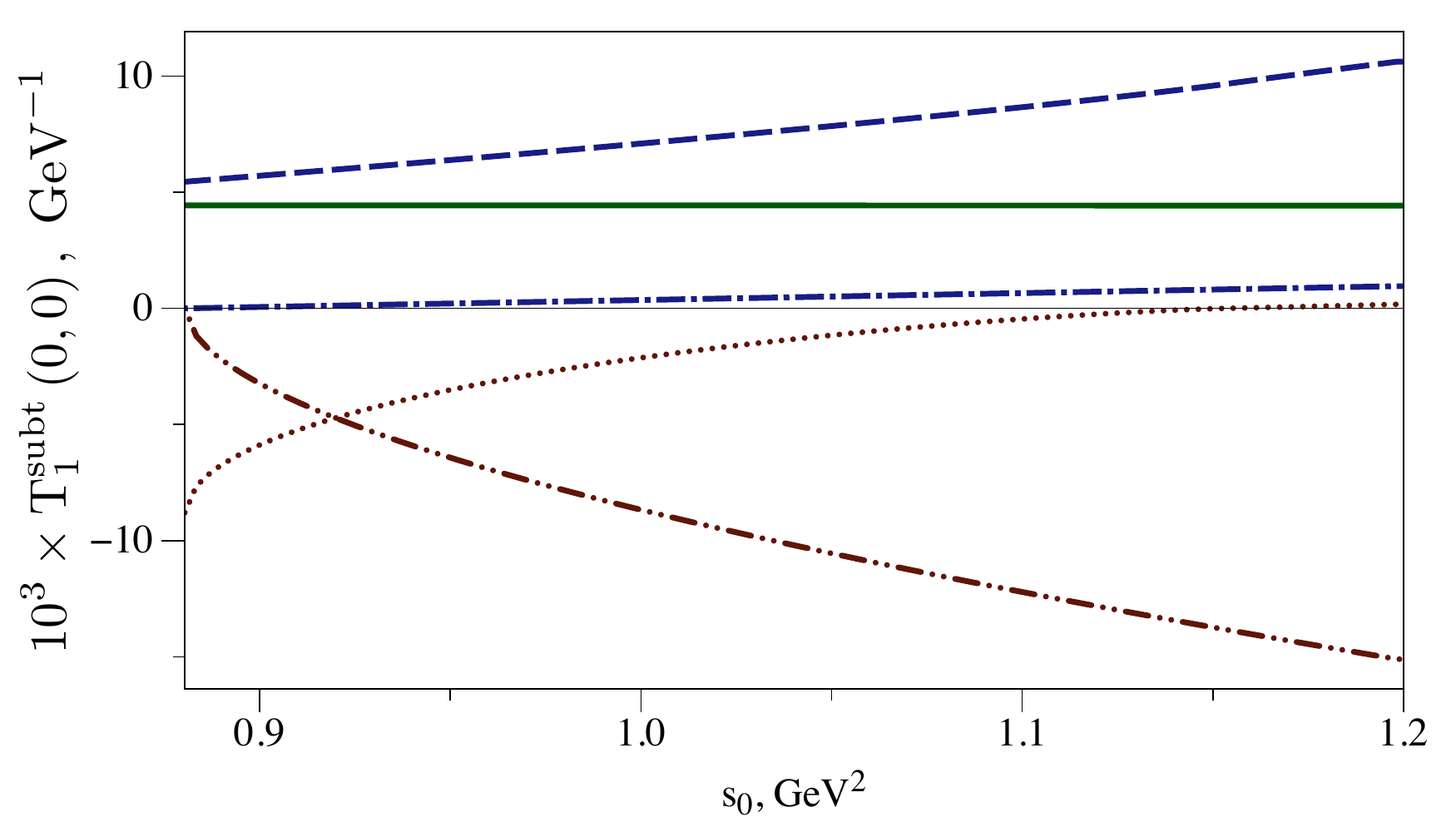}}
\caption{The contribution of the individual terms in Eq. (\ref{subtraction_from_HE}) to ${\mathrm{T}}^\mathrm{subt}_1(0,0)$ as function of $ s_0 $. Dashed curve: the dispersion integral contribution from the BC fit $ \sim \int_{W^2_{\mathrm{thr}}}^{10~ \mathrm{GeV}^2} (F^{\mathrm{BC}}_1 - F_1^{\mathrm{R}}) $. Dashed-dotted curve: the dispersion integral contribution from the DL fit $ \sim \int_{10~\mathrm{GeV}^2}^\infty (F^{\mathrm{DL}}_1 - F_1^{\mathrm{R}}) $. Dotted curve:  the dispersion integral contribution $ \sim - \int^{W^2_{\mathrm{thr}}}_{s_0} F_1^{\mathrm{R}} $ due to $F_1^{\mathrm{R}}$. 
Dashed double-dotted curve: the contribution from the real part $ T_1^R(0,0) $ according to Eq.~(\ref{T10Q2}). 
Solid curve: sum of all terms in Eq. (\ref{subtraction_from_HE}), yielding the $s_0$-independent value of $T^{\mathrm{subt}}_1 (0,0) $.}
\label{s0_independence}
\end{figure}

We furthermore provide another consistency check of our numerical implementation. As the Regge function $\mathrm{T}_1^\mathrm{R}$ 
of Eq.~(\ref{regge_T1}) has an arbitrary scale $\tilde \nu_0$ (or equivalently $s_0$), the total result should not depend 
on the specific choice of this parameter. We demonstrate this in Fig. \ref{s0_independence}, where we illustrate how 
the $s_0$ dependence of the individual contributions in Eq. (\ref{subtraction_from_HE}) adds up to yield the total result 
which is independent of $s_0$.  

In Fig.~\ref{experimental_T1} we present the empirically extracted subtraction function $ \mathrm{T}^{\mathrm{subt}}_1 ( 0, \tilde{Q}^2 ) $ of Eq. (\ref{subtraction_from_HE}) and compare it with the subtraction functions of Refs.~\cite{Birse:2012eb,Alarcon:2013cba}. The subtraction function $ \mathrm{T}^{\mathrm{subt}}_1(0, \tilde Q^2)$ should vanish linearly when $\tilde Q^2 \to 0$ according to Eq.~(\ref{floop}). This general property therefore provides a quality check on the accuracy of an empirical determination as described above. 
One notices from Fig.~\ref{experimental_T1} that the value of  $ \mathrm{T}^{\mathrm{subt}}_1$ at $\tilde Q^2 = 0$ is compatible with zero within $ 1-1.5 ~\sigma$. We would like to notice, however, that at present such empirical determination can unfortunately only give the correct order of magnitude of 
 $ \mathrm{T}^{\mathrm{subt}}_1 ( 0, \tilde{Q}^2 ) $. This is partly due to the non-perfect match between the proton $F_1$ fits for the resonance region and the large $W$ region, as we have shown in Fig.~\ref{F1_plots}. Despite this caveat, it seems, however, that with increasing $ \tilde{Q}^2 $, $ \mathrm{T}^{\mathrm{subt}}_1 ( 0, \tilde{Q}^2 ) $ changes sign in the range somewhere between 0.1 and 0.4 GeV$^2$, which may be an indication of the range up to which the ChPT-based results can be used. To provide a more accurate determination of the functional dependence of $ \mathrm{T}^{\mathrm{subt}}_1 ( 0, \tilde{Q}^2 ) $, a combined fit of all proton $F_1$ structure function data over the whole range of $ W $, incorporating the Regge behavior at large $W$ would be desirable. At intermediate values of $Q^2$, below and around 1 GeV$^2$, this will also require one to have more accurate data in the intermediate $W$ range between $3 - 10$~GeV. In the lower end of this range, such data can be provided by the JLab 12 GeV facility.   

Using our empirical determination of $ \mathrm{T}^{\mathrm{subt}}_1 ( 0, \tilde{Q}^2 ) $, we can extract $\beta ( \tilde{Q}^2) $ dividing $ \mathrm{T}^{\mathrm{subt}}_1(0,\tilde Q^2)$ by $\tilde Q^2$ according to Eq.~(\ref{floop}). For the purpose of combining our empirical estimate of $ \mathrm{T}^{\mathrm{subt}}_1(0, \tilde Q^2)$ with the empirical value of $\beta(0)$ as determined from RCS, we use the central curve in the empirically determined error band of $ \mathrm{T}^{\mathrm{subt}}_1 ( 0, \tilde{Q}^2 )$ (green band in Fig.~\ref{experimental_T1}) to extract $\beta(\tilde Q^2)$  in the range $  \tilde Q^2 > 0.12~\mathrm{GeV}^2$, and extrapolate it by a linear function to the PDG value of $\beta_M$ at $\tilde{Q}^2 = 0 $. The resulting curve is displayed in Fig.~\ref{beta_Q2}. We will use the latter curve in the following to provide an empirical estimate for the subtraction function contribution to the TPE correction for the muon-proton elastic scattering at small momentum transfer.

\subsection{TPE correction from the subtraction function}
\label{sec4d}

Using Eqs.~(\ref{TPE_correction} - \ref{OPETPE}), we can now estimate the TPE correction due to the $ \mathrm{T}^{\mathrm{subt}}_1 ( 0, \tilde{Q}^2 )$ contribution to the first term in the hadronic tensor 
of Eq.~(\ref{vvcs_tensor}).
Performing the traces in Eq.~(\ref{OPETPE}) explicitly, 
the subtraction function results in the following TPE correction in the region of low momentum transfers: 
\ber \label{TPE_subtraction_all_muon}
 \delta^{\mathrm{subt}}_{2 \gamma} & = &  \frac{32 \pi G_E }{\varepsilon G^2_E + \tau G^2_M }  \frac{1-\varepsilon}{1-\varepsilon_0} \frac{1}{M} \Re \mathop{\mathlarger{\int}} \frac{  i \mathrm{d}^4 \tilde{q}}{\left( 2 \pi \right)^4} \beta \left(\tilde{Q}^2 - \frac{Q^2}{4} \right)   \Pi_{K}^{+} \Pi_{K}^{-}\Pi_{Q}^{+} \Pi_{Q}^{-} \nonumber \\
&\times& \left \{ \left(K \cdot P \right)  m^2 \left(\tilde{Q}^2 - \frac{Q^2}{4}\right)^2 
+ \frac{1}{2} \left( Q^2 \left( P\cdot\tilde{q} \right) - 4 \left(K\cdot P\right) \left( K\cdot\tilde{q} \right)  \right) \left(K \cdot \tilde{q}\right)  \left(\tilde{Q}^2 +\frac{Q^2}{4}\right) \right \}, \nonumber \\
\eer
where the lepton (photon) propagators $\Pi^\pm_K$ ($\Pi^\pm_Q$) are defined as in Eq.~(\ref{eq:propdef}). 
The second term within the curly brackets of Eq. (\ref{TPE_subtraction_all_muon}) can be simplified to yield the 
expression
\ber \label{TPE_subtraction_all_evaluate}
\delta^{\mathrm{subt}}_{2 \gamma} & = & \frac{32 \pi m^2 G_E }{\varepsilon  G^2_E + \tau G^2_M } \frac{1-\varepsilon}{1-\varepsilon_0} \frac{\left( K \cdot P \right)}{M}  \nonumber \\
&\times& \Re\mathop{\mathlarger{\int}} \frac{ i \mathrm{d}^4 \tilde{q}}{\left( 2 \pi \right)^4} \beta \left(\tilde{Q}^2 - \frac{Q^2}{4} \right)   \Pi_{K}^{+} \Pi_{K}^{-}\Pi_{Q}^{+} \Pi_{Q}^{-} 
  \left \{ \left(\tilde{Q}^2 - \frac{Q^2}{4}\right)^2 - \frac{ 2 \left(K \cdot\tilde{q}\right)^2} {m^2 + \frac{Q^2}{4}} \left(\tilde{Q}^2 +  \frac{Q^2}{4}\right) \right\}, \nonumber \\
\eer
making explicit the overall proportionality of $ \delta^{\mathrm{subt}}_{2 \gamma} $ to the squared lepton mass $ m^2 $.

\begin{figure}[h]
\centering{\includegraphics[width=0.7\textwidth]{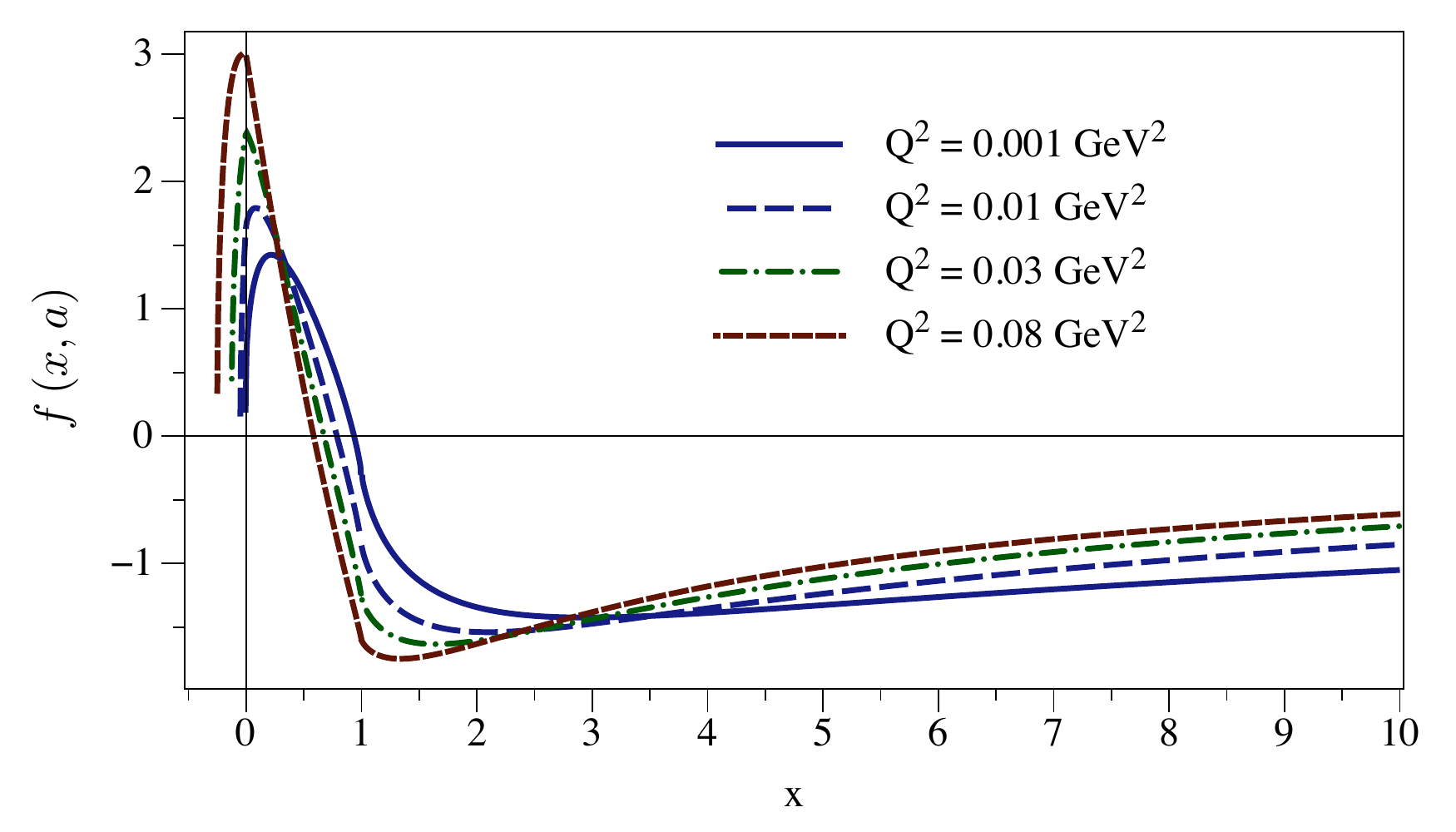}}
\caption{The weighting function $ f$ of Eq. (\ref{weighting_function_expression})   for the range of 
$Q^2$ values of the MUSE experiment.} 
\label{weighting_function_plot}
\end{figure}

The integration in Eq. (\ref{TPE_subtraction_all_evaluate}) is performed through a Wick rotation, as detailed in Appendix \ref{subtraction_term}, and the resulting TPE correction is given by
\ber \label{TPE_subtraction_all_integration_result}
&& \delta^{\mathrm{subt}}_{2 \gamma} = \frac{m^2 G_E\left(Q^2\right) }{\varepsilon  G^2_E\left(Q^2\right) + \tau G^2_M\left(Q^2\right) } \frac{1-\varepsilon}{1-\varepsilon_0} \frac{\left( K \cdot P \right)}{M} \frac{Q}{K} \mathop{\mathlarger{\int}} \limits^{~~\infty}_{x_{\mathrm{min}}}  f \left( x, a \right) \beta \left(\frac{Q^2 \left(x -1 \right)}{4}\right) \frac{\mathrm{d} x}{4 \pi},
\eer
in terms of the dimensionless variable $ x = 4 \tilde{Q}^2/Q^2 $ with the weighting function $ f \left( x, a \right) $:
\ber \label{weighting_function_expression}
f \left( x, a \right) & = &  -  \frac{2 x^{\Theta \left( 1-x \right)}}{\sqrt{1 + a}} \Theta \left(x \right)  + \frac{1+ 2 a - x}{ 1 + a} \frac{ | 1 - x |}{1+x} \left\{ \ln \left| \frac{  x- z  }{ x + z } \right| \Theta \left(x \right) \Theta \left(1-x \right) \right. \nonumber \\
&& \left. +   \ln \left |   \frac{z + 1 }{ z - 1} \right | \Theta \left(x - 1\right)  +  \ln \left| \frac{  x- z  }{ x + z }  \frac{z - 1 }{ z + 1} \right| \Theta \left(x - x_{\mathrm{min}} \right) \Theta \left(-x \right) \right\} ,
\eer
with
\ber \label{weighting_function_notation}
z   =   \frac{1 - x - \sqrt{\left(1+x\right)^2 + 4 a x} }{2 \sqrt{ 1 + a }}, \quad
 x_{\mathrm{min}} = - \left(  \sqrt{1+a} - \sqrt{a} \right)^2, ~\quad a  = \frac{4 m^2}{Q^2}. 
\eer
At small momentum transfers the result of Eq. (\ref{TPE_subtraction_all_integration_result}) starts from a term proportional to $ Q^2 $. We show the $ x $ (or $ \tilde{Q}^2 $) dependence of the weighting function of Eq. (\ref{weighting_function_expression}) in Fig. \ref{weighting_function_plot}. 

The TPE contribution due to the subtraction function also provides a correction to the 2S - 2P muonic hydrogen Lamb shift, which is the largest hadronic uncertainty in this precise quantity \cite{Pohl:2010zza, Antognini:1900ns}.  
Using the ChPT-based results for $ \beta ( \tilde{Q}^2 ) $ as input, this TPE correction was 
estimated in Refs.~\cite{Birse:2012eb,Alarcon:2013cba} and found to be too small to resolve the proton radius puzzle. The correction from the above discussed empirically determined subtraction function to the 2S energy level in muonic hydrogen, after integration up to $ \tilde{Q}^2 = 1~\mathrm{GeV}^2 $, yields:\footnote{Note that for the total TPE correction to the muonic hydrogen 2S level one needs to add to the subtraction function contribution also the dispersive contribution, which was evaluated based on data in Ref.~\cite{Carlson:2011zd}. }
\ber \label{lamb_shift_estimate}
\Delta E^{\mathrm{subt}}_{\mathrm{\mathrm{2S}}} \approx 2.3 ~\mu \mathrm{eV},
\eer
which is in fair agreement with the estimate of Birse et al. \cite{Birse:2012eb}, though slightly smaller: $ \Delta E^{\mathrm{subt}}_{\mathrm{\mathrm{2S}}} \approx 4.2 \pm 1.0 ~\mu \mathrm{eV} $. Our result of Eq. (\ref{lamb_shift_estimate}) is also within errors of the analogous evaluation of Ref. \cite{Gorchtein:2013yga}, where the authors assumed the existence of $ J = 0 $ fixed pole. It was speculated in Ref. \cite{Miller:2012ne} that to explain the proton radius puzzle would require a huge enhancement of $ \beta ( \tilde{Q}^2 ) $ at large $ \tilde{Q}^2 $. In order to account for 
the experimentally observed discrepancy in $\Delta E_{\mathrm{2S}}$ of around 310 $\mu$eV~\cite{Carlson:2015jba}, it would require an  around two orders of magnitude larger TPE correction than the naturally expected result from the ChPT estimates. 
For this purpose, an ad hoc subtraction function, proposed to be added as an extra contribution on top of the ChPT-based  subtraction functions discussed above, was conjectured in Ref. \cite{Miller:2012ne} with the following functional form:
\ber
\beta_{\mathrm{extra}} ( \tilde{Q}^2 ) =   \left( \frac{\tilde{Q}^2}{M^2_0} \right)^2 \frac{\beta_M}{ \left( 1 + \tilde{Q}^2 / \Lambda_0^2 \right)^5}, \qquad M_0 = 0.5 ~\mathrm{GeV}, \quad \Lambda_0 = 3.92 ~\mathrm{GeV}.
\eer
In such a scenario, the large $ \tilde{Q}^2 $ region would also dominate the TPE correction to the muon-proton elastic scattering, and the integral of Eq.~(\ref{TPE_subtraction_all_integration_result}) would be approximated by 
\ber \label{TPE_subtraction_approximation}
\delta^{\mathrm{subt}}_{2 \gamma,0} \approx - \frac{3 m^2 G_E }{\varepsilon  G^2_E + \tau G^2_M }  \frac{1-\varepsilon}{1-\varepsilon_0}  \frac{\left( K \cdot P \right)}{ \pi M}  \mathop{\mathlarger{\int}} \limits^{~\infty}_0 \beta (\tilde{Q}^2 ) \frac{\mathrm{d} \tilde{Q}^2}{ \tilde{Q}^2} \approx  - \frac{ 3 Q^2 m^2 }{2 \pi E } \mathop{\mathlarger{\int}} \limits^{~\infty}_0  \beta (\tilde{Q}^2) \frac{\mathrm{d} \tilde{Q}^2}{\tilde{Q}^2},
\eer
where the last step gives the approximate expression in the limit $ Q^2 \ll M^2, ~M E, ~E^2 $. This approximation corresponds in magnitude with the result of Ref. \cite{Miller:2012ne} for $ \mu^{-} p $ scattering, however, it differs by an overall sign.

\begin{figure}[h]
\centering{\includegraphics[width=.7\textwidth]{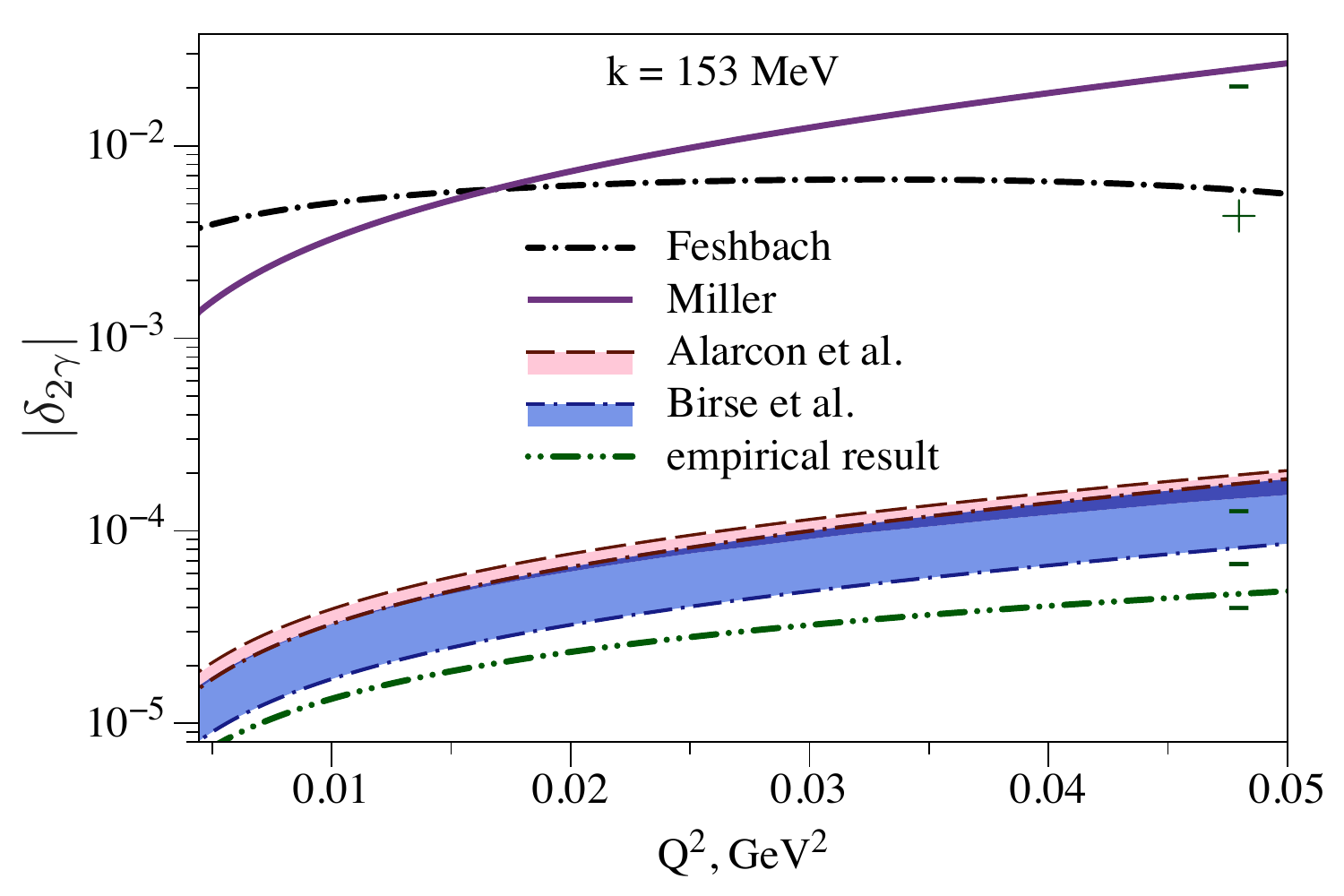}}
\caption{Subtraction function contribution to the TPE correction in elastic muon-proton scattering for the muon {\it lab} momentum $ \mathrm{k} = 153 ~\mathrm{MeV} $. Blue band: result for the HBChPT-based subtraction function~\cite{Birse:2012eb}. Pink band: result for the BChPT-based subtraction function~\cite{Alarcon:2013cba}. Dashed doubly-dotted (green) curve: result based on the empirical subtraction function, corresponding with the dashed doubly-dotted curve in Fig.~\ref{beta_Q2}. Solid curve: result based on the conjectured subtraction function of Ref.~\cite{Miller:2012ne}. 
The (black) dashed-dotted curve is the Feshbach term of Eq. (\ref{full_feshbach}) for a pointlike Dirac particle corrected by the recoil factor $(1 + m / M)$. The sign labels on the curve show the sign of the corresponding expressions for $ \mu^- p $ scattering.}
\label{delta_TPE_vs_Q2}
\end{figure}

\begin{figure}[h]
\centering{\includegraphics[width=0.7\textwidth]{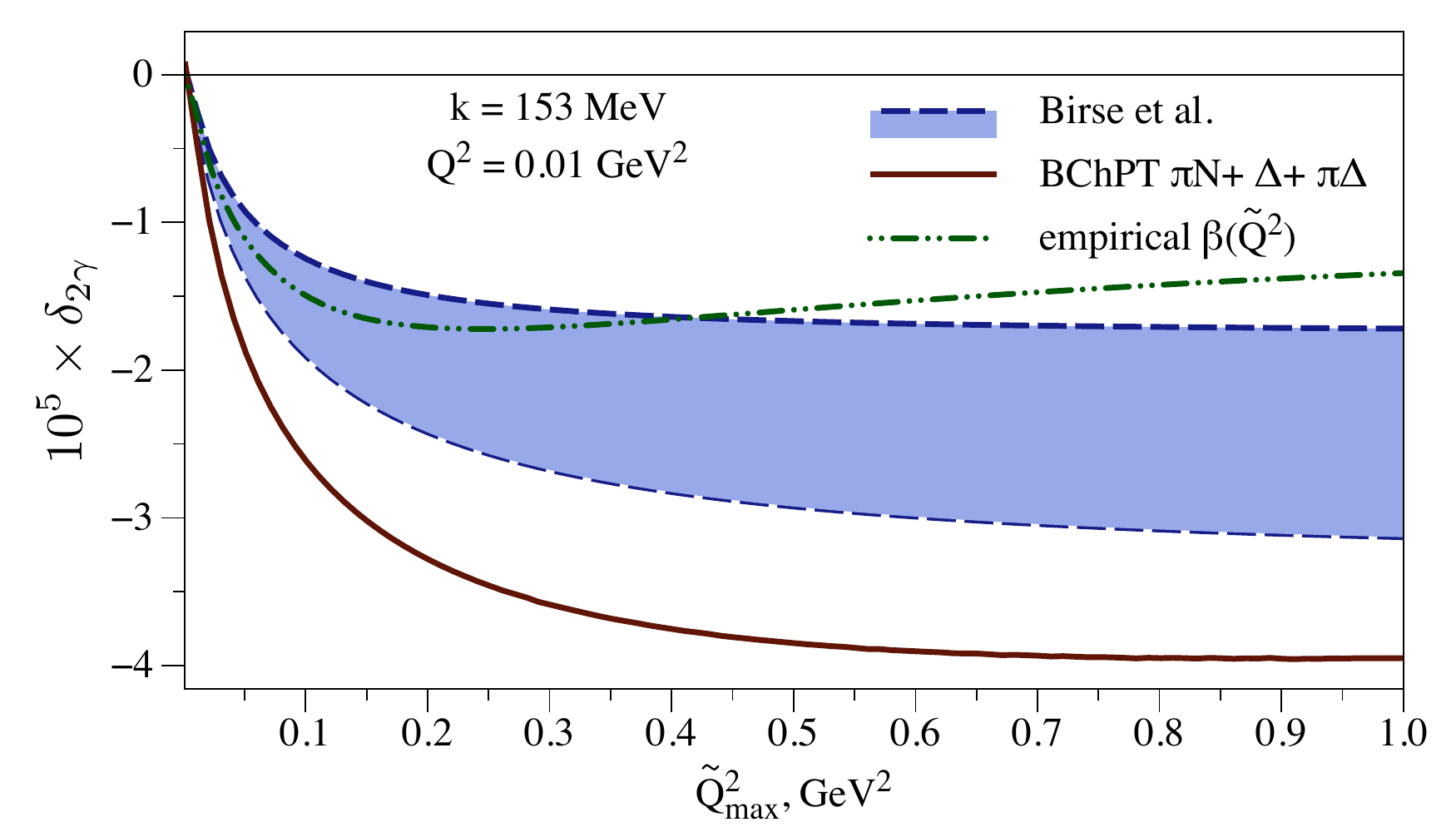}}
\caption{The dependence of the integral of Eq. (\ref{TPE_subtraction_all_integration_result}) on the upper integration limit $ \tilde{Q}^2_{\mathrm{max}} $ for three different estimates of the subtraction function $\beta(\tilde Q^2)$ as described in the text.} 
\label{beta_saturation}
\end{figure}

In Fig.~\ref{delta_TPE_vs_Q2} we compare the TPE correction to elastic muon-proton scattering (for MUSE kinematics) due to the above discussed ChPT as well as  empirically determined subtraction functions. To estimate the size of uncertainties of the BChPT result~\cite{Alarcon:2013cba}, we plot a band corresponding with a variation of the upper integration limit in Eq. (\ref{TPE_subtraction_all_integration_result}) between $ \tilde{Q}^2 = 0.9 - 5$~GeV$^2$. We notice that the HBChPT and BChPT results are in agreement within their uncertainties. 
The TPE correction due to the empirically extracted subtraction function is also shown on Fig.~\ref{delta_TPE_vs_Q2}, giving a similar though slightly smaller result. This can be understood as the empirically determined $\beta(\tilde Q^2)$ changes sign as function of $\tilde Q^2$. The region of $\tilde Q^2$ contributing to the above result is shown in Fig.~\ref{beta_saturation}. One sees that the TPE integral has largely converged for an upper integration limit value of around $\tilde Q^2_\mathrm{max} \sim 1$~GeV$^2$. 

In Fig.~\ref{delta_TPE_vs_Q2}, we furthermore also show the TPE correction to elastic muon-proton scattering resulting from the subtraction function conjectured in Ref.~\cite{Miller:2012ne} to explain the proton radius puzzle through enhancing the 
TPE corrections by nearly two orders of magnitude. Even though the weighting functions entering the TPE corrections in the muonic hydrogen Lamb shift and the elastic muon-proton scattering are different, one notices from Fig.~\ref{delta_TPE_vs_Q2} that the subtraction function of Ref.~\cite{Miller:2012ne} also yields a nearly two order of magnitude larger TPE correction for the elastic muon-proton scattering.   
To put this in perspective, we also display in Fig.~\ref{delta_TPE_vs_Q2} the model independent estimate of the elastic TPE contribution, which has to be added on top of the inelastic TPE contribution, and which is due to the Feshbach term of Eq. (\ref{full_feshbach}) corrected by the recoil factor $(1 + m / M)$. One notices that the use of such large subtraction function 
would yield an inelastic TPE correction to elastic muon-proton scattering which in magnitude already would exceed the elastic Feshbach contribution around $Q^2 = 0.02$~GeV$^2$, and would increase further with increasing $Q^2$.  

\section{Inelastic contribution to TPE correction}
\label{sec5}

Besides the subtraction function contribution, the inelastic TPE correction to elastic muon-proton scattering includes the contribution of the DR integrals in Eqs.~(\ref{forward_VVCS_DR_T1}) and (\ref{forward_VVCS_DR_T2}). 
Using Eqs.~(\ref{TPE_correction}) and (\ref{OPE}) and working out the traces in Eq.~(\ref{OPETPE}), the corresponding contributions from the unpolarized proton structure functions $ F_1 $ and $ F_2 $ to $\delta_{2 \gamma}$ are given by
\ber
 \label{TPE_inelastic1} \delta_{2 \gamma}^{F_1} & = & F ~ \Re \mathop{\mathlarger{\int}} \limits^{~~ \infty}_{W^2_{\mathrm{thr}}} \mathrm{d} W^2   \mathop{\mathlarger{\int}} \frac{ i \mathrm{d}^4 \tilde{q}}{\left( 2 \pi \right)^4}  \Pi_{Q}^{+} \Pi_{Q}^{-} F_1\left(W^2,\tilde{Q}^2-\frac{Q^2}{4}\right) \frac{(P\cdot\tilde{q})^2  }{ W^2 - P^2 + \tilde{Q}^2  }  \nonumber \\
&& \hspace{4cm}\times  \frac{\left\{  A \left( \Pi_{K}^{-} + \Pi_{K}^{+} \right)  + B   \left( \Pi_{K}^{-} - \Pi_{K}^{+} \right) \right\}}{\left( \left( P + \tilde{q} \right)^2 - W^2  \right) \left( \left( P - \tilde{q} \right)^2 - W^2 \right)} , \\
 \label{TPE_inelastic2} \delta_{2 \gamma}^{F_2}& = & F ~ \Re \mathop{\mathlarger{\int}} \limits^{~~ \infty}_{W^2_{\mathrm{thr}}} \mathrm{d} W^2  \mathop{\mathlarger{\int}} \frac{ i \mathrm{d}^4 \tilde{q}}{\left( 2 \pi \right)^4}  \Pi_{Q}^{+} \Pi_{Q}^{-}  F_2\left(W^2,\tilde{Q}^2-\frac{Q^2}{4}\right) \nonumber \\
&& \hspace{4cm}\times \frac{  \left\{  C \left( \Pi_{K}^{-} + \Pi_{K}^{+} \right)  + D   \left( \Pi_{K}^{-} - \Pi_{K}^{+} \right) \right\}}{\left( \left( P + \tilde{q} \right)^2 - W^2 \right) \left( \left( P - \tilde{q} \right)^2 - W^2 \right) } , 
\eer
with definitions introduced in Eqs.~(\ref{eq:defeps} - \ref{eq:propdef}) and the following notation:
\ber 
F &=& \frac{8 e^2}{M^2}  \frac{G_E}{\varepsilon G_E^2 + \tau G^2_M} \frac{1-\varepsilon}{1-\varepsilon_0}, \nonumber \\
 A & = & -4 m^2\left(K\cdot P\right), \nonumber \\
 B &= & \frac{\tilde{Q}^2+\frac{Q^2}{4}}{\tilde{Q}^2-\frac{Q^2}{4}}\left( Q^2 \left( P\cdot\tilde{q} \right) - 4 \left(K\cdot P\right) \left( K\cdot\tilde{q} \right)  \right), \nonumber \\
C & = & \frac{1}{2} \left(K\cdot P\right) \left( 4 \left(K\cdot P\right)^2 - Q^2  P^2 \right) +  \frac{Q^2}{\tilde{Q}^2-\frac{Q^2}{4}}  \left(P\cdot\tilde{q}\right)  \left(\left(K\cdot\tilde{q} \right) P^2 - \left(P\cdot\tilde{q}\right) \left(K\cdot P\right) \right) , \nonumber\\
D & = & -\frac{1}{4} \left( P^2 + \frac{ (P\cdot\tilde{q})^2 Q^2}{\left(\tilde{Q}^2-\frac{Q^2}{4}\right)^2 }\right) \left(  Q^2 \left(P\cdot\tilde{q}\right) - 4 \left(K\cdot P\right) \left(K\cdot\tilde{q} \right)    \right) \nonumber \\
&-& \frac{1}{2} \left(P\cdot\tilde{q}\right) \frac{ \tilde{Q}^2-\frac{3 Q^2}{4} }{ \tilde{Q}^2-\frac{Q^2}{4}  } \left(4 \left(K\cdot P\right)^2  - Q^2  P^2  \right).
 \label{abc}
\eer

Our numerical studies of the inelastic TPE contribution indicate that, in the limit $ Q^2 \ll m^2, ~M^2, ~ M E $, the momentum transfer expansion starts with a $ Q^2 $ term and contains no $ Q^2 \ln Q^2 $ type of non-analyticity . This is unlike the elastic electron-proton scattering case, where in the limit $ m^2 \ll Q^2 \ll M^2 , ~ M E $ a non-analytic behavior of the type 
$ Q^2 \ln Q^2$ is present at low $Q^2$~\cite{Brown:1970te, Tomalak:2015aoa}. 

To provide numerical estimates of the inelastic TPE contribution to elastic muon-proton scattering due to the dispersion integrals, we express the corresponding integrals of Eqs.~(\ref{TPE_inelastic1}, \ref{TPE_inelastic2}) in the form
\ber \label{TPE_integrand}
\delta_{2 \gamma}^{F_1, F_2}  = \delta_{2 \gamma}^{F_1} + \delta_{2 \gamma}^{F_2} =\mathop{\mathlarger{\int}} \limits^{~~\infty}_{W^2_{\mathrm{thr}}} f \left( W \right) \mathrm{d} W^2.
\eer

\begin{figure}[h]
\centering{\includegraphics[width=0.5\textwidth]{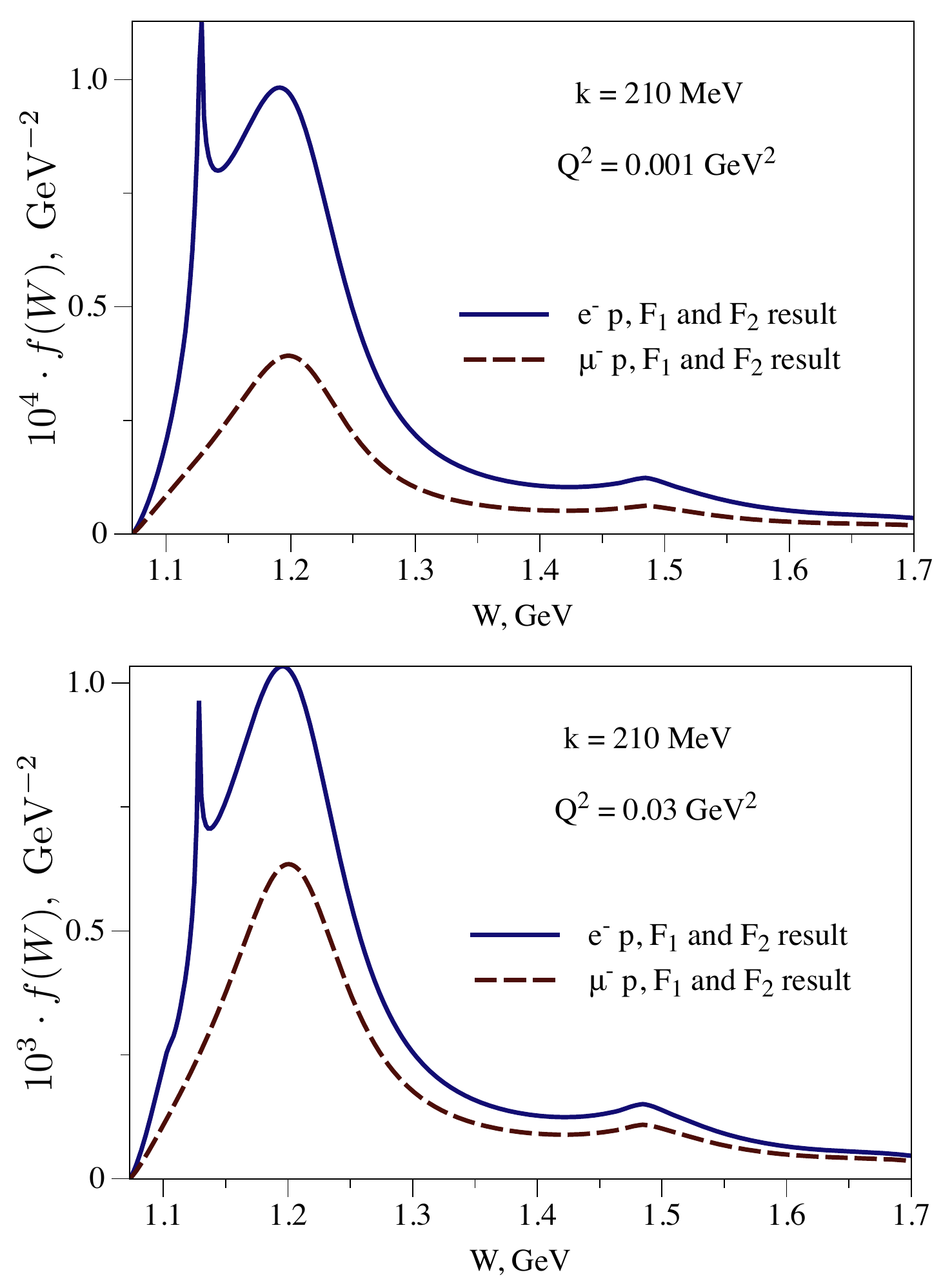}}
\caption{$ W$ dependence of the integrand $f \left( W \right)$ which determines the inelastic TPE correction, as given by Eq. (\ref{TPE_integrand}). The integrand is shown for the case of $ e^{-} p $ and $ \mu^{-} p $ elastic scattering. The external kinematics (indicated on the plots) correspond with the MUSE experiment. }
\label{W_integrand1}
\end{figure}

\begin{figure}[h]
\centering{\includegraphics[width=0.5\textwidth]{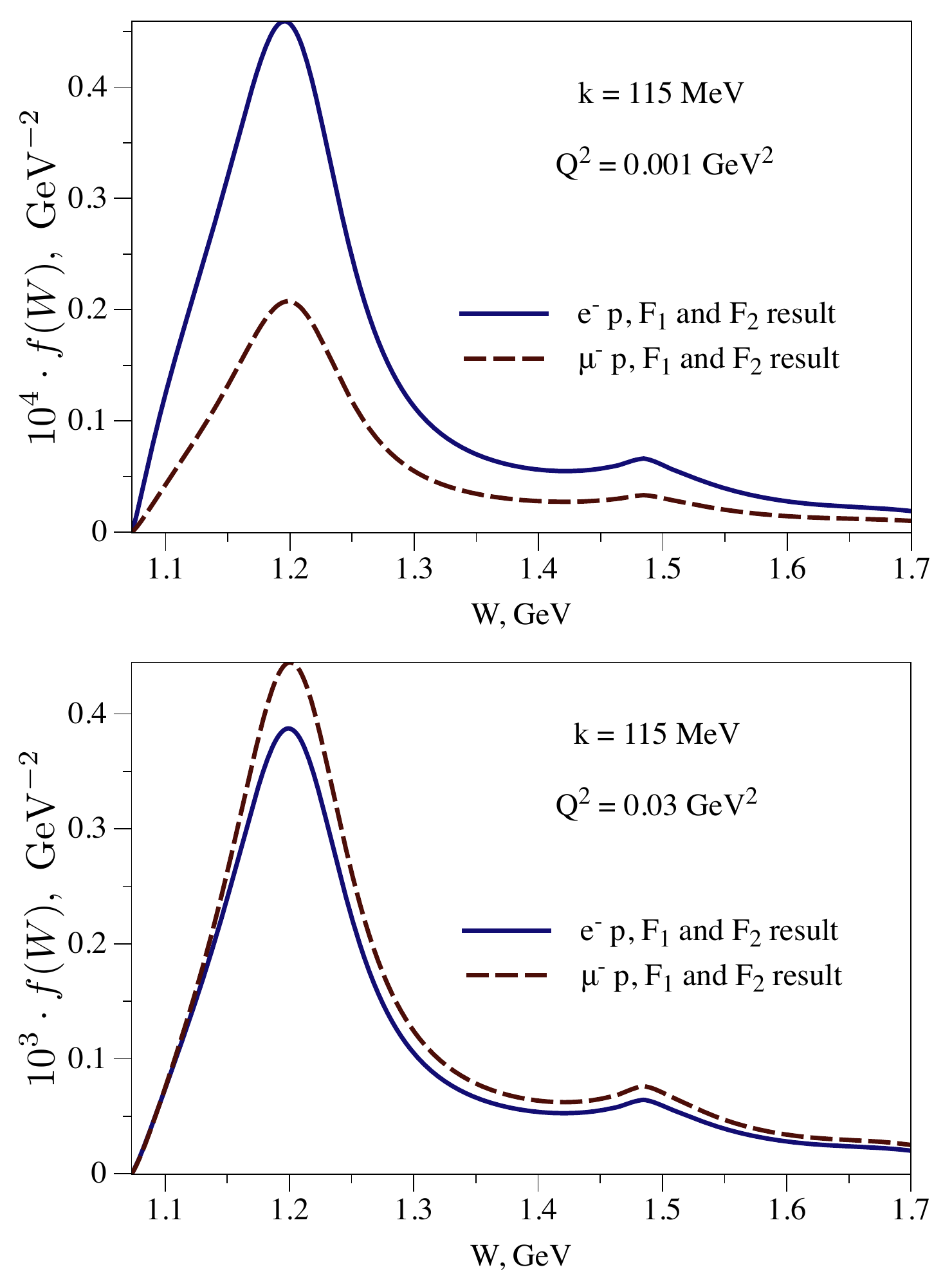}}
\caption{Same as Fig. \ref{W_integrand1}, but for the lepton momentum $ \mathrm{k} = 115 ~\mathrm{MeV} $.}
\label{W_integrand2}
\end{figure}

In Figs. \ref{W_integrand1}, \ref{W_integrand2} we compare the $ W$ dependence of the integrand $ f \left( W \right) $ in 
Eq.~(\ref{TPE_integrand}) for the $\mu^- p$ and $e^- p$ elastic scattering processes. As input for the 
proton structure functions $F_1$ and $F_2$, we use the fit performed by Christy and Bosted~\cite{Christy:2007ve}. 
We find that the $\tilde Q^2$ integrations are well saturated when performed up to $\tilde Q^2 = 8$~GeV$^2$, which is the largest value covered by the BC fit. As a test, we extended the BC fit beyond its fit region and found that the relative contribution from the region $ 8~\mathrm{GeV}^2 < \tilde{Q}^2 < 12~\mathrm{GeV}^2 $ is smaller than $ 0.015 ~\% $. 

Figures \ref{W_integrand1}, \ref{W_integrand2}  show results in different kinematics corresponding with the MUSE experiment. 
The TPE corrections to $e^- p $ are sizeably larger than for the $\mu^- p $ case at low $Q^2$. With increasing 
$Q^2$, the $\mu^- p$ TPE corrections increase, as is evident from the result at lower beam momentum in Fig.~\ref{W_integrand2}, where at $Q^2 = 0.03$~GeV$^2$ both corrections reach similar sizes.  
We furthermore notice in Fig.~\ref{W_integrand1} that the integrand for the elastic $e^- p$ scattering displays a narrow peak corresponding with the quasi-real photon singularity (for both photons), see Ref.~\cite{Tomalak:2015aoa}, which is absent for the $\mu^- p$ case.

To estimate the inelastic TPE correction to elastic lepton-proton scattering, we find that the $W$ integration in Eq.~(\ref{TPE_integrand}) is well saturated when performed up to 3.1 GeV, which is the largest value covered by the BC fit. 
When again extending the BC fit beyond its fit range, for the purpose of a test, we checked that the 
relative contribution from the region $ 3.1 ~\mathrm{GeV} < W < 4 ~\mathrm{GeV} $ to 
$\delta_{2 \gamma}^{F_1, F_2}$ is smaller than $ 1.5 ~\% $. We estimate the uncertainties of the numerical integration coming  from the integration regions outside the BC fit and from the inaccuracies in the BC fit at $ 5 - 6 ~\% $ level.

\begin{figure}
\centering{\includegraphics[width=\textwidth]{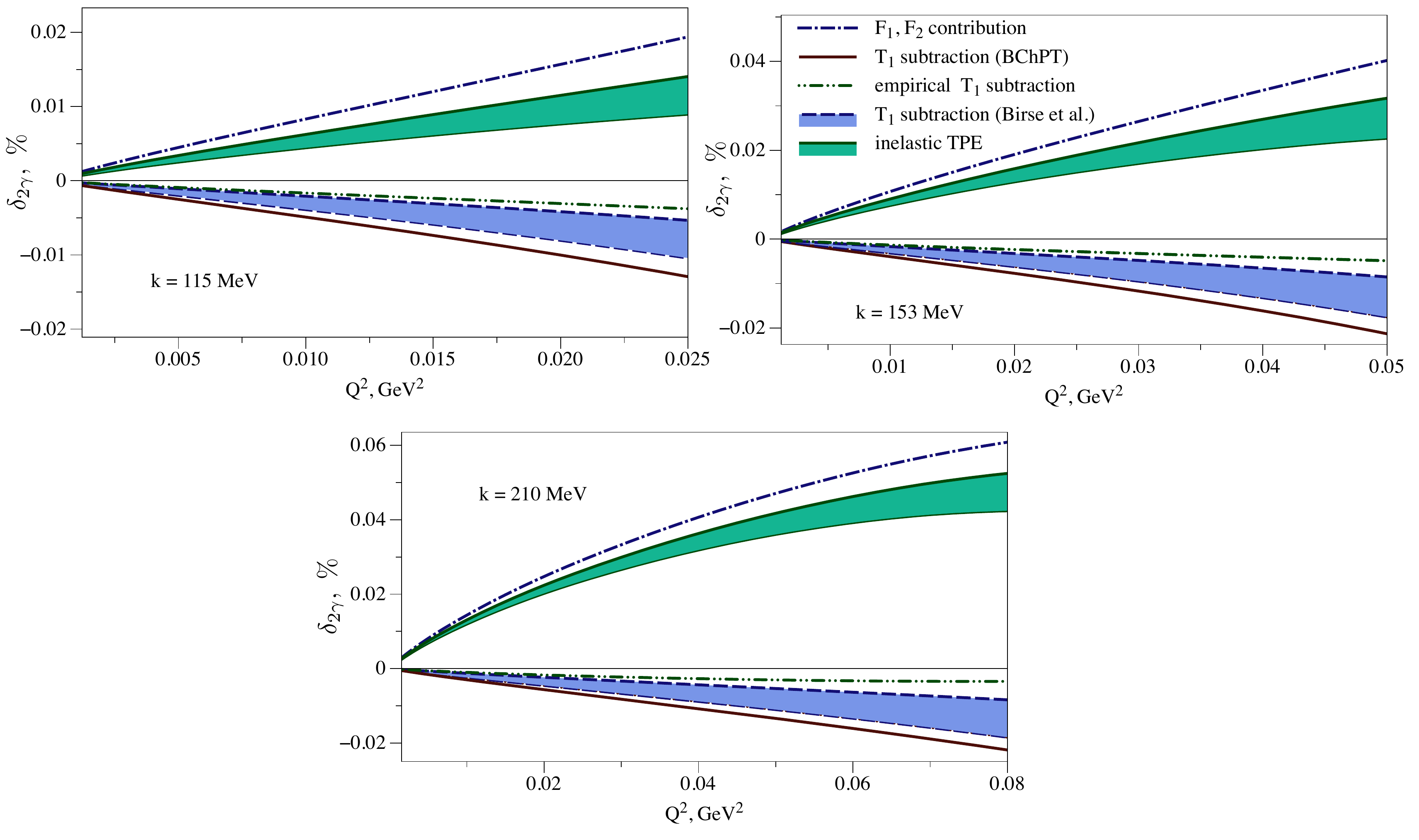}}
\caption{TPE correction for $ \mu^{-} p $ elastic scattering for three different muon {\it lab} momenta as planned in the MUSE experiment. The TPE correction due to the subtraction function is shown for three subtraction function inputs: Birse et al. \cite{Birse:2012eb} (blue bands),  BChPT~\cite{Alarcon:2013cba} (solid curves), and the empirical determination as described in Section~\ref{sec4} (dashed doubly-dotted curves).  The inelastic TPE correction due to the dispersion integrals over the proton structure functions $ F_1$ and $ F_2 $ is shown by the dashed-dotted curves. The resulting total inelastic TPE correction (sum of both) is shown by the green bands using the subtraction function of Birse et al.}
\label{muon_results}
\end{figure}

The resulting inelastic TPE corrections for the elastic $\mu^- p$ scattering process are shown in Fig.  \ref{muon_results} 
as a function of $Q^2$ for three values of muon beam momentum, corresponding with the MUSE kinematics. 
Note that the muon beam {\it lab} momenta $ \mathrm{k} = 115 ~\mathrm{MeV} $, $ \mathrm{k} = 153 ~\mathrm{MeV} $, and $ \mathrm{k} = 210 ~\mathrm{MeV} $, correspond with the kinematically allowed regions of  $ Q^2 < 0.039~\mathrm{GeV}^2$, $ Q^2 < 0.066~\mathrm{GeV}^2$, and $ Q^2 < 0.116~\mathrm{GeV}^2$, respectively. We notice that for the small momentum transfers corresponding with the MUSE kinematics, the inelastic TPE corrections to elastic $\mu^- p$ scattering are very small, in the range of $\delta_{2 \gamma} \sim 5 \times 10^{-4}$. This is well below the anticipated cross section precision of around 1 \% of the MUSE experiment. Furthermore, we notice that the TPE corrections due to the subtraction function and the dispersive $F_1, F_2$ structure function integrals come with opposite signs, leading to a partial cancellation.

\begin{figure}
\centering{\includegraphics[width=0.6\textwidth]{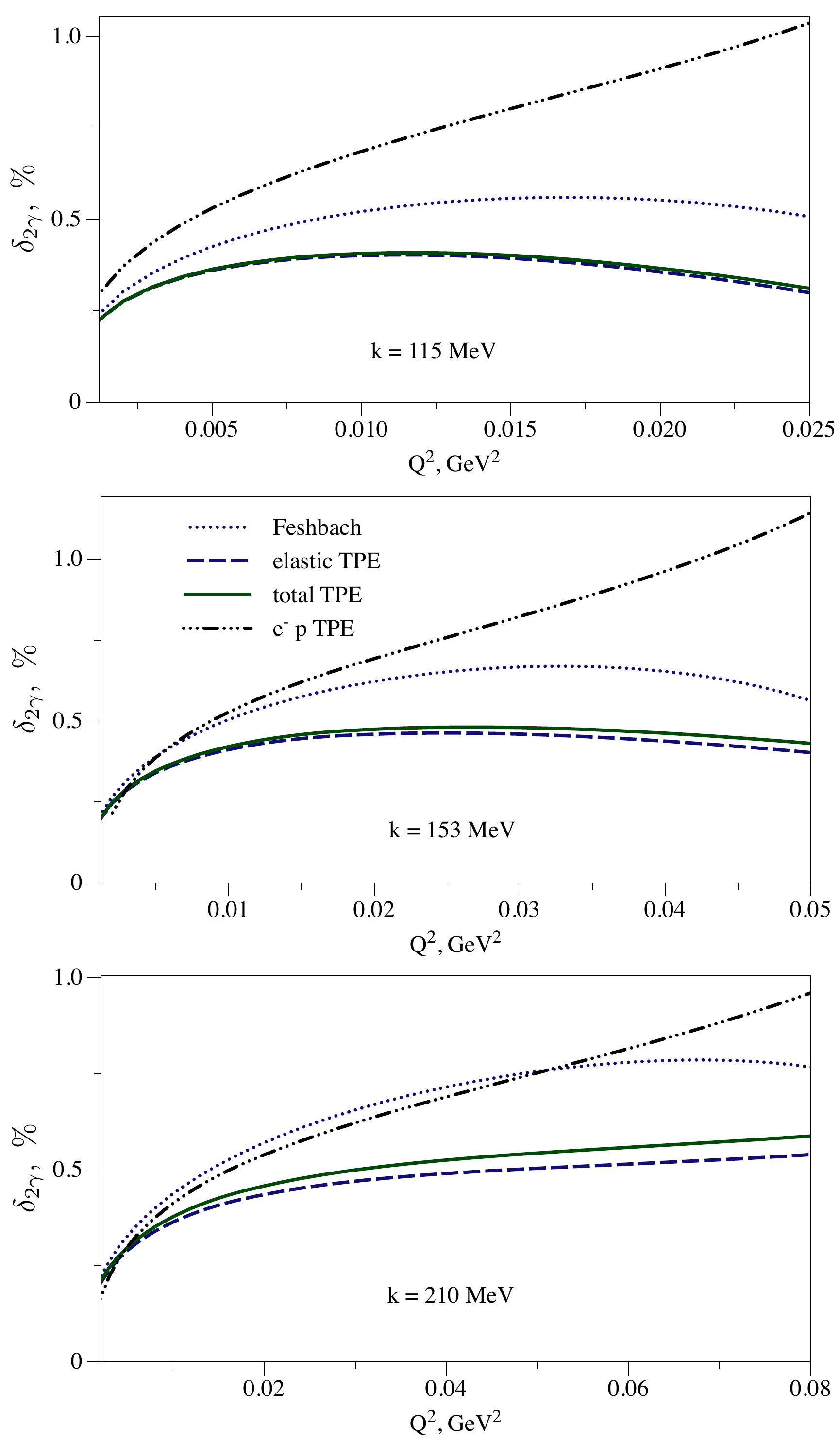}}
\caption{The total TPE correction for $\mu^- p$ elastic scattering is shown as sum of the elastic TPE,  the TPE correction from the $ F_1$ and $ F_2 $ proton structure functions and the TPE correction from the subtraction function of Ref. \cite{Birse:2012eb}. It is compared with the Feshbach term for point-like particles, see Eq. (\ref{full_feshbach}), corrected by the recoil factor $ (1 + m / M )$, the elastic contribution based on the box graph evaluation with dipole form factors~\cite{Tomalak:2014dja} and the $ e^- p $ total TPE correction \cite{Tomalak:2015aoa}.}
\label{muon_results2}
\end{figure}

We present in Fig.  \ref{muon_results2} the total TPE correction as a sum of the Born TPE correction of Ref.~\cite{Tomalak:2014dja}, corresponding with a proton intermediate state, and the inelastic TPE of this work using the subtraction function of Birse et al.~\cite{Birse:2012eb}. We compare our result with the Feshbach term of Eq.~(\ref{full_feshbach}) for a point-like Dirac particle corrected by the recoil factor $ (1 + m / M )$, with the elastic TPE correction based on the box graph evaluation with proton form factors of the dipole form \cite{Tomalak:2014dja},  and with the corresponding TPE correction for elastic $ e^- p $ scattering of Ref.~\cite{Tomalak:2015aoa}. Contrary to the electron-proton scattering case, where the subtraction function contribution is negligible \cite{Tomalak:2015aoa} as it is proportional to the lepton mass squared, in the case of muon-proton scattering the inelastic proton structure function contribution is partially canceled by the $ \mathrm{T}_1 $ subtraction function resulting in a negligibly small inelastic TPE correction for the MUSE kinematics. Only with increased lepton beam energy or when going to larger $Q^2$ values one needs to start accounting for the inelastic TPE correction, which shifts the total correction a little closer to the Feshbach result.

\section{Conclusions}
\label{sec6}

In this work we have estimated the TPE correction to muon-proton elastic scattering at low momentum transfer. 
For the elastic (proton) intermediate state contribution, we have derived a low-momentum transfer expansion accounting for all terms due to the non-zero lepton mass. Besides the elastic contribution, we have accounted for the inelastic intermediate states by expressing the TPE process at low momentum transfer approximately through the forward doubly virtual Compton scattering. 
The input in our evaluation of the inelastic TPE correction is given by the unpolarized proton structure functions and by one subtraction function, corresponding with the forward Compton amplitude $\mathrm{T}_1$ at zero photon energy. For the latter, we have 
compared two estimates based on heavy-baryon and baryon chiral perturbation theory with an empirical determination. 
For the empirical determination, we have expressed the subtraction function through an unsubtracted dispersion relation for 
the amplitude $\mathrm{T}_1 - \mathrm{T}_1^R$. The function $\mathrm{T}_1^R$ is suitably defined through a Regge pole fit such that at high-energies   $(\mathrm{T}_1 - \mathrm{T}_1^R) \to 0$, ensuring convergence and applicability of the unsubtracted dispersion relation. We have provided a numerical evaluation of the subtraction function 
based on a Regge fit of high-energy proton structure function data. It was found that the extracted subtraction function is compatible in magnitude with the chiral perturbation theory calculations, and thus cannot explain the proton radius puzzle through missing TPE corrections, which would have required a total TPE correction which is larger by around an order of magnitude compared with the empirical and chiral perturbation theory-based evaluations.  
Besides the subtraction function, the second part of the inelastic TPE contribution was obtained through dispersion integrals  over the unpolarized proton structure functions. For the latter, we used a fit of the data in the proton resonance region. Using our formalism, we have provided estimates for the total TPE corrections in the kinematics of forthcoming muon-proton elastic scattering data of the MUSE experiment. We found that in the MUSE kinematics, the elastic TPE contribution largely dominates, and the size of the inelastic TPE contributions is within the anticipated error of the forthcoming data.

\section*{Acknowledgements}
We thank Nikolay Kivel and Vladyslav Pauk for useful discussions, Vadim Lensky and Vladimir Pascalutsa for providing us with the chiral perturbation theory results on the proton magnetic polarizability and Carl Carlson for providing us with the parametrizations of the unpolarized proton structure functions. This work was supported in part by the Deutsche Forschungsgemeinschaft DFG in part through the Collaborative Research Center [The Low-Energy Frontier of the Standard Model (SFB 1044)], in part through the Graduate School [Symmetry Breaking in Fundamental Interactions (DFG/GRK 1581)], and in part through the Cluster of Excellence [Precision Physics, Fundamental Interactions and Structure of Matter (PRISMA)].
 
\appendix

\section{Elastic TPE contribution at small $ Q^2 $}
\label{lowQ2limit}

In this Appendix we study the low momentum transfer limit of the TPE correction due to the proton intermediate state. The leading terms in the momentum transfer ($ Q^2 $) expansion arising from the proton intermediate state contribution (elastic TPE) are given by the graphs with two point Dirac couplings $ \gamma_\mu $ in the lower blob of the diagrams in Fig. \ref{TPE_kinematics}. The unpolarized part of the doubly virtual Compton scattering (VVCS) tensor for a Dirac point particle results in the TPE correction $ \delta^{\mathrm{QED}}_{2 \gamma} $ given by
\ber \label{lp_elastic_integral}
\delta^{\mathrm{QED}}_{2 \gamma} & = & \frac{ 32 \left(K\cdot P\right) e^2}{M^2 \left( \varepsilon + \tau \right)} \frac{1-\varepsilon}{1-\varepsilon_0} \Re \mathop{\mathlarger{\int}} \frac{  i \mathrm{d}^4 \tilde{q}}{\left( 2 \pi \right)^4} \Pi_{P}^{+} \Pi_{P}^{-}\Pi_{K}^{+} \Pi_{K}^{-}\Pi_{Q}^{+} \Pi_{Q}^{-} \nonumber \\
&& \left \{  \left(2 \tilde{q}^2 + \frac{3}{2} Q^2\right) \left(K\cdot P\right) \left(K\cdot \tilde{q}\right) \left(P\cdot\tilde{q}\right) \right.  \nonumber \\
&& - \left. \left(\tilde{q}^2 + \frac{Q^2}{4}\right) \left(  \left(P^2 + \frac{Q^2}{4} \right)\left(K\cdot\tilde{q}\right)^2 +  \left(K^2 + \frac{Q^2}{4} \right) \left(P\cdot\tilde{q}\right)^2\right)    \right.  \nonumber \\
&& - \left. \left(\tilde{q}^2 + \frac{Q^2}{4}\right)^2 \left( \left(K \cdot P \right)^2 - \frac{Q^2}{4} \left( M^2 + m^2 \right) +  \frac{ \left( q \cdot \tilde{q} \right)^2  - Q^2 \tilde{q}^2}{8}  \right) \right. \nonumber \\
&& \left. + \frac{Q^2}{2 \left( P \cdot K \right)}  \left(K\cdot\tilde{q}\right) \left(P\cdot\tilde{q}\right) \left(  \left(P\cdot\tilde{q}\right)^2 +  \left(K\cdot\tilde{q}\right)^2 \right) - 2 \left(K\cdot\tilde{q}\right)^2 \left(P\cdot\tilde{q}\right)^2 \right. \nonumber \\
&& \left. + \frac{\left(P^2 + K^2 \right) Q^2}{4 \left( K \cdot P \right)}   \left(K\cdot \tilde{q}\right) \left(P\cdot\tilde{q}\right) \left( \tilde{q}^2 - \frac{3}{4} Q^2 \right) \right. \nonumber \\
&& \left.  -  \frac{\left(K\cdot \tilde{q}\right) \left(P\cdot\tilde{q}\right)}{2 \left(K\cdot P \right) }  \left( \left( q \cdot \tilde{q} \right)^2 \frac{Q^2}{4}  + Q^2 \left(  \frac{5 \tilde{q}^2 Q^2}{8} -\frac{\tilde{q}^4}{4} - \frac{17}{64} Q^4 \right) \right) \right \},
\eer
with $ \varepsilon $ the photon polarization parameter:
\ber
\varepsilon = \frac{E^2 - 2 M E \tau  - M^2 \tau}{E^2 - 2 M E \tau + M^2 \tau + 2 M^2 \tau^2 - m^2 \left(1 + \tau\right)},
\label{eq:defeps}
\eer
and where we used the kinematic notations:
\ber
\tau = \frac{Q^2}{4 M^2}, \qquad \varepsilon_0 = \frac{2 m^2}{Q^2}.
\label{eq:deftau}
\eer
We have introduced the averaged lepton (proton) four-momenta $ K (P) $ by
\ber
K = \frac{1}{2} \left( k + k' \right), \quad P = \frac{1}{2} \left( p + p' \right),
\label{eq:defkin}
\eer
and the propagator notation
\ber
 \Pi_{P}^{\pm} = \frac{1}{\left(P\pm \tilde{q}\right)^2 - M^2}, ~~~~~\Pi_{K}^{\pm} = \frac{1}{\left(K\pm \tilde{q}\right)^2 - m^2},~~~~~\Pi_{Q}^{\pm} = \frac{1}{\left(\tilde{q} \pm q/2\right)^2 - \mu^2},
 \label{eq:propdef}
\eer
where the photon mass $ \mu $ plays the role of IR regulator.

When studying now the low-$Q^2$ expansion $ Q^2 \ll m^2, ~M^2, ~M^2 {\rm k}^2 /s$, of the expression of Eq. (\ref{lp_elastic_integral}) we find the analog of the Feshbach term $ \delta_F $ \cite{McKinley:1948zz}, the IR divergent piece $ \delta^{\mathrm{IR}}_{2 \gamma} $, and a logarithmic correction:
\ber \label{TPE_expansion_massive}
\delta^{\mathrm{QED}}_{2 \gamma}  & \to & \delta^{\mathrm{IR}}_{2 \gamma} + \delta_F   + \frac{\alpha Q^2}{2 \pi M E} \ln  \frac{Q^2}{2 M E} \left( 1 + \frac{2 E }{{\rm k} }  \ln \frac{{\rm k} - E + m}{{\rm k}+E-m} \right) + \mathrm{O} \left(\frac{Q^2}{M^2}, \frac{Q^2}{m^2},  \frac{s Q^2}{M^2 {\rm k}^2} \right), \nonumber \\\label{elastic_TPE}\\
\delta_F & \to & \alpha \pi \frac{Q}{2E} \frac{M+m}{M},  \label{feshbach_limit} \\ 
 \delta^{\mathrm{IR}}_{2 \gamma} & \to & \frac{\alpha E Q^2 }{ \pi M {\rm k}^2}  \ln \frac{\mu^2}{Q^2} \left( 1 + \frac{m^2}{E {\rm k}}   \ln  \frac{{\rm k} - E + m}{{\rm k}+E-m}  \right) \label{IR_piece},
\eer
with the lepton momentum in the lab frame $ {\rm k} \equiv | \vec{k} | $.

We also provide the more general expansion of Eq. (\ref{lp_elastic_integral}) in the low-$Q^2$ limit $ Q^2 \ll M^2, ~M^2 {\rm k}^2 /s $, where $ Q^2 $ needs not be very small relative to the squared lepton mass. For such expansion, the leading $ Q^2 $ terms are given by
\ber \label{elastic_TPE_general}
\delta_{2 \gamma}^{\mathrm{QED}} & \to & \delta^{\mathrm{IR}}_{2 \gamma} + \frac{ \alpha \pi Q}{2 E} +   \frac{ \alpha E Q^2}{2 \pi M {\rm k}^2 } \left(  \frac{\frac{Q^2}{4}}{m^2 + \frac{Q^2}{4}} + \frac{2 {\rm k}}{E} \ln  \frac{{\rm k} - E + m}{{\rm k}+E-m} \right) \ln \frac{Q^2}{2 M E}  \nonumber \\
&& + \frac{16 \pi \alpha Q^2 }{{\rm k}^2 M E} \frac{  E^2  \frac{Q^4}{8} + m^2 E^2 Q^2 + m^4 {\rm k}^2 }{m^2 + \frac{Q^2}{4}} C \left(m,Q^2\right) +  \mathrm{O} \left(\frac{Q^2}{M^2}, \frac{s Q^2}{M^2 {\rm k}^2} \right),
\eer
with
\ber
C\left(m,Q^2\right) & = & \frac{1}{16 \pi^2} \frac{1}{Q^2} \frac{1}{\sqrt{1+\frac{4m^2}{Q^2}}} \left \{ \ln \frac{Q^2}{m^2} \ln \left(\frac{1+\sqrt{1+\frac{4m^2}{Q^2}}}{-1+\sqrt{1+\frac{4m^2}{Q^2}}} \right) - \mathrm{Li}_2 \left(\frac{2}{1+\sqrt{1+\frac{4m^2}{Q^2}}} \right)  \right. \nonumber \\
&& \left. - \mathrm{Li}_2 \left( \frac{1-\sqrt{1+\frac{4m^2}{Q^2}}}{2} \right) -\frac{1}{2} \ln^2 \left(\frac{2}{-1+\sqrt{1+\frac{4m^2}{Q^2}}} \right)+ \frac{5}{6} \pi^2 \right \},
\eer
where $ \mathrm{Li}_2(x) $ denotes the dilogarithm function. The leading IR divergent piece is given by Eq. (\ref{IR_piece}). 
In the limit $ Q^2 \ll m^2 $, the result of Eq. (\ref{elastic_TPE_general}) reduces to the expression of Eq. (\ref{elastic_TPE}). When taking the massless limit $  m^2 \ll Q^2 $ as limiting case of Eq. (\ref{elastic_TPE_general}), we also recover the expressions of Refs. \cite{Brown:1970te,Tomalak:2015aoa}.

\section{Regge poles residues of the proton structure function $F_1$ from high-energy data}
\label{experimental_T1_details}

\begin{figure}[h]
\centering{\includegraphics[width=.75\textwidth]{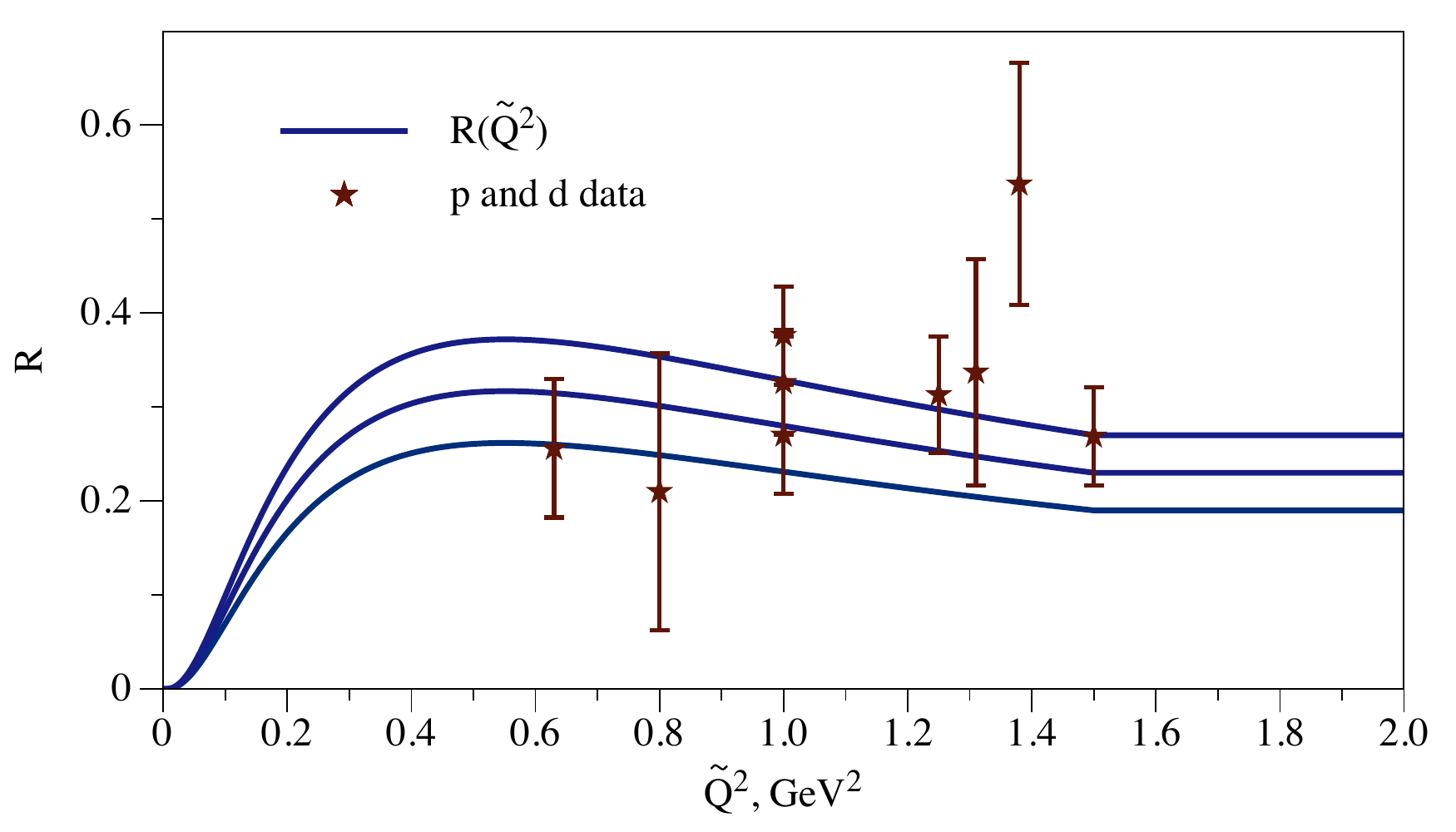}}
\caption{$\tilde Q^2$ dependence of the ratio $ R = \sigma^{\gamma p}_L / \sigma^{\gamma p}_T $. The experimental result $ R_0 = 0.23 \pm 0.04 $ in the region $ \tilde{Q}^2 > 1.5 ~\mathrm{GeV}^2 $ from H1 and ZEUS~\cite{Andreev:2013vha} is connected with the ratio taken from the BC fit~\cite{Christy:2007ve} (central curve). The error band reflects the experimental uncertainty in the value of $R$ from the fit to the H1 and ZEUS data. The data points are from Refs.~\cite{Whitlow:1990gk,Dasu:1993vk,Arneodo:1996qe}.}
\label{ratio}
\end{figure}

The high-energy limit ($\tilde{\nu}$ very large at fixed $\tilde Q^2$) of the proton structure function $F_1$ is often parameterized through a Regge pole fit as
\ber \label{regge_pole_residue}
F_1  ( \tilde{\nu}, \tilde{Q}^2 ) & \underset{\tilde{\nu}\gg}{ \longrightarrow} & \sum_{{\alpha_0} \ge 0} \gamma_{\alpha_0} (\tilde{Q}^2 )  \tilde{\nu}^{{\alpha_0}},
\eer
where $ \gamma_{\alpha_0} (\tilde{Q}^2 ) $ are the leading Regge poles residues, which 
can be extracted from the high-energy inclusive electron-proton scattering data. We will determine these residues from 
the Donnachie-Landshoff (DL) fit~\cite{Donnachie:2004pi} to data 
for the proton structure function $F_2$ in the region of very small Bjorken variable $ x_{\mathrm{Bj}} \equiv \tilde{Q}^2 / \left( 2 M \tilde{\nu} \right) $:
\ber
F_2 \left(\tilde \nu = \frac{\tilde Q^2}{2 M x_{\mathrm{Bj}}}, \tilde{Q}^2 \right)  \underset{x_{\mathrm{Bj}} \ll 1}{ \longrightarrow}  f_0 ( \tilde{Q}^2 ) x_{\mathrm{Bj}}^{-\varepsilon_0} + f_1 ( \tilde{Q}^2 ) x_{\mathrm{Bj}}^{-\varepsilon_1}  + f_2 ( \tilde{Q}^2 ) x_{\mathrm{Bj}}^{-\varepsilon_2}, 
\eer
where 
\ber
 f_0 ( \tilde{Q}^2 ) & = & A_0 \left( \frac{\tilde{Q}^2}{1 + \tilde{Q}^2 / Q^2_0 } \right)^{1 + \varepsilon_0} \left( 1 + \tilde{Q}^2 / Q^2_0 \right)^{\varepsilon_0 / 2}, \\
 f_1 ( \tilde{Q}^2 ) & = & A_1 \left( \frac{\tilde{Q}^2 }{1 + \tilde{Q}^2 / Q^2_1} \right)^{1 + \varepsilon_1}, \\
 f_2 ( \tilde{Q}^2 ) & = & A_2 \left( \frac{\tilde{Q}^2 }{1 + \tilde{Q}^2 / Q^2_2 } \right)^{1 + \varepsilon_2}, 
\eer
with parameters values (using $ \mathrm{GeV} $ units for all mass scales)\cite{Donnachie:2004pi}:
\ber
A_0 & = & 0.00151, \qquad ~~A_1 = 0.658, \qquad ~~~~~A_2 = 1.01, \nonumber \\
Q^2_0 & = & 7.85, \quad ~~~~~~~~~~Q^2_1 = 0.6 , \quad ~~~~~~~~~~~ Q^2_2 = 0.214 , \nonumber \\
\varepsilon_0 & = & 0.452, \qquad ~~~~~~ \varepsilon_1 = 0.0667, \qquad ~~~~~\varepsilon_2 = -0.476.
\eer
The $ F_1 $ structure function in the high-energy region is then obtained:
\ber
F_1 ( \tilde \nu, \tilde{Q}^2 )  \underset{\tilde{\nu}\gg}{ \longrightarrow}  \frac{M \tilde \nu}{\tilde{Q}^2}  \frac{F_2 ( \tilde \nu, \tilde{Q}^2 ) }{1 + R },
\label{eq:F1fromF2}
\eer
where $ R \equiv \sigma^{\gamma p}_L/\sigma^{\gamma p}_T$  is the ratio of longitudinal to transverse virtual photon absorption cross sections on a proton. 
We will use the experimental result $ R_0 = 0.23 \pm 0.04 $ at $ \tilde{Q}^2 > 1.5 ~\mathrm{GeV}^2 $ from the H1 and ZEUS collaborations \cite{Andreev:2013vha}, and approximate $R$ in our numerical estimates by the following expression, independent of $W^2 \equiv 2 M \tilde \nu + M^2 - \tilde Q^2$:
\ber
R = R (\tilde{Q}^2 ) = R_0 \Theta \left( \tilde{Q}^2 - 1.5 ~\mathrm{GeV}^2 \right) + R_{\mathrm{BC}} (\tilde{Q}^2 )  \Theta \left( - \tilde{Q}^2 + 1.5 ~\mathrm{GeV}^2 \right),
\label{eq:ratioR}
\eer
where $ R_{\mathrm{BC}} (\tilde{Q}^2 ) $ is value obtained in the Christy and Bosted fit~\cite{Christy:2007ve} 
evaluated at $ W^2 \approx 2.63 ~ \mathrm{GeV}^2 $. The latter corresponds with the $ W^2 $ value for which the ratio $R$ from the BC fit $ R_{\mathrm{BC}} (\tilde{Q}^2 = 1.5 ~\mathrm{GeV}^2) \approx 0.23 $, and thus goes over into the H1/ZEUS value at $\tilde Q^2 > 1.5$~GeV$^2$. We use the relative uncertainties from the data of Ref. \cite{Andreev:2013vha} in the whole $ \tilde{Q}^2 $ region. 
We show the resulting functional form of $ R (\tilde{Q}^2) $  
in Fig. \ref{ratio}, and we compare its value with the data from Refs.~\cite{Whitlow:1990gk,Dasu:1993vk,Arneodo:1996qe} in the range 
$\tilde Q^2 < 1.5$~GeV$^2$. We notice that our 
parameterization of $R$ yields good agreement with the data.  

Adopting the above Regge parameterization for $F_2$, with the ratio $R$ from Eq.~(\ref{eq:ratioR}), we obtain from Eq.~(\ref{eq:F1fromF2}) for $F_1$ the following Regge pole residues entering Eq.~(\ref{regge_pole_residue}):
\ber
\gamma_{1+\varepsilon_i} (\tilde{Q}^2 )  = \frac{1}{2} \, \frac{f_i(\tilde{Q}^2 ) }{1 +  R ( \tilde{Q}^2 )} \left( \frac{2 M}{ \tilde{Q}^2} \right )^{1+\varepsilon_i}.
\eer

\section{Subtraction function TPE correction: evaluation of integrals}
\label{subtraction_term}

The first integral in the subtraction function TPE correction of Eq. (\ref{TPE_subtraction_all_evaluate}) is of the type:
\ber \label{subtraction_1_initial}
\hspace{-0.5 cm} I_1 & = &  \mathop{\mathlarger{\int}} \frac{ i \mathrm{d}^4 \tilde{q}}{\left( 2 \pi \right)^4} g ( \tilde{Q}^2 ) \Pi_{Q}^{+} \Pi_{Q}^{-} \Pi_{K}^{+} \Pi_{K}^{-},
\eer
where $ g ( \tilde{Q}^2 ) $ is given by
\ber
g ( \tilde{Q}^2 ) = \left( \tilde{Q}^2 - \frac{Q^2}{4} \right)^2 \beta \left( \tilde{Q}^2 - \frac{Q^2}{4}\right).
\eer
 Accounting for the symmetry  $ \tilde{q} \to - \tilde{q} $, the integral $ I_1 $ can be expressed as
\ber \label{subtraction_1_initial_2}
I_1 = \frac{1}{2} \mathop{\mathlarger{\int}} \frac{  i \mathrm{d}^4 \tilde{q}}{\left( 2 \pi \right)^4} \frac{g ( \tilde{Q}^2 ) \Pi_K^+  \left( \Pi_Q^+ +  \Pi_Q^- \right)}{\left( \tilde{Q}^2 -  \frac{Q^2}{4} \right) \left( \tilde{Q}^2 +\frac{Q^2}{4} \right)} .
\eer
The azimuthal angle integration is trivial, the polar angle integration gives the same result for both terms of Eq. (\ref{subtraction_1_initial_2}), such that the integral $ I_1 $ can be written as
\ber \label{subtraction_1_initial_3}
I_1 = \mathop{\mathlarger{\int}} \frac{  i \mathrm{d}^4 \tilde{q}}{\left( 2 \pi \right)^4}\frac{g ( \tilde{Q}^2 ) \Pi_K^+   \Pi_Q^+ }{\left( \tilde{Q}^2 -  \frac{Q^2}{4} \right) \left( \tilde{Q}^2 +\frac{Q^2}{4} \right)} .
\eer

We evaluate the integral conveniently in the lepton Breit frame defined by
\ber
K = K \left(1,0,0,0 \right),   \qquad  q = Q \left(0,0,0,1 \right),
\eer
and perform the integral through a Wick rotation. The integration contour crosses the lepton propagator poles in this frame during the Wick rotation, as detailed in Ref. \cite{Tomalak:2015aoa}. The integral of Eq. (\ref{subtraction_1_initial_3}) is given by the sum of the integral along the imaginary axis, which we denote by $ I_1^W $, and the lepton pole contribution, which we denote by $ I_1^p $. 

The integral along the imaginary axis  $ I^W_1$ can be evaluated by the Gegenbauer polynomial technique, see Appendix B of Ref. \cite{Tomalak:2015aoa} for some technical details. It results in an integral over the dimensionless variable $ x \equiv 4 \tilde{Q}^2 / Q^2 $ as
\ber \label{I1_Wick}
I^W_1 = \frac{1}{Q^4 \sqrt{1+a}} \mathop{\mathlarger{\int}} \limits_{0}^{~~\infty} \frac{ \mathrm{d} x}{2 \pi^2} \frac{ g \left( \frac{x Q^2}{4} \right)}{ | 1 - x^2 | }  \ln \left |   \frac{ z  + 1 }{ z - 1}  \right |  ,
\eer
with the notation of Eq. (\ref{weighting_function_notation}). 

The contribution of the pole $ \tilde{q}_0 = K - \sqrt{\tilde{q}^2 + m^2} + i \varepsilon  $ which is enclosed by the Wick rotation contour for the integral $ I_1 $ is given by
\ber \label{I1_pole}
I^p_1 =  \frac{1}{Q^4 \sqrt{1+a}}  \mathop{\mathlarger{\int}} \limits^{~~1}_{x_{\mathrm{min}}} \frac{\mathrm{d} x }{ 2 \pi^2}  \frac{ g \left( \frac{x Q^2}{4} \right)}{ 1 - x^2 }  \ln \left | \frac{ \left( z - 1 \right) \left( x- z \right) }{ \left( z + 1 \right) \left( x + z \right)} \right | \Theta \left(x  + \left(  \sqrt{1+a} - \sqrt{a} \right)^2 \right),
\eer
with $ a $ defined in Eq. (\ref{weighting_function_notation}). Note that the lower integration limit in Eq. (\ref{I1_pole}) is given by
\ber 
x_{\mathrm{min}} = - \left( \sqrt{1+a} - \sqrt{a} \right)^2,
\eer
which has limits $ x_{\mathrm{min}} \to -1 $ for $ m^2 \ll Q^2 $, and $ x_{\mathrm{min}} \to - \frac{Q^2}{16 m^2} $ for $ Q^2 \ll m^2 $.

Subsequently, we evaluate the second integral in Eq. (\ref{TPE_subtraction_all_evaluate}):
\ber \label{subtraction_2_initial}
I_2 & = & -  \int \frac{ i \mathrm{d}^4 \tilde{q}}{\left( 2 \pi \right)^4} \frac{2 \left( K\cdot\tilde{q} \right)^2}{K^2} \left(\tilde{Q}^2 +  \frac{Q^2}{4}\right) \beta \left( \tilde{Q}^2 - \frac{Q^2}{4}\right) \Pi_{K}^{+} \Pi_{K}^{-}\Pi_{Q}^{+} \Pi_{Q}^{-}. 
\eer
Performing similar steps as for the $ I_1 $ integral we obtain
\ber \label{subtraction_3_initial}
I_2 =   \int \frac{  i \mathrm{d}^4 \tilde{q}}{\left( 2 \pi \right)^4} \frac{\beta \left( \tilde{Q}^2 - \frac{Q^2}{4}\right)}{2 K^2} \left \{ - \left( \tilde{Q}^2 - \frac{Q^2}{4} \right)  \Pi_K^+ \Pi_Q^+ + \left( \tilde{Q}^2 + \frac{Q^2}{4} \right)\Pi_Q^+ \Pi_Q^-   \right \}. 
\eer
The integral from the first term can be obtained from the $ I_1 $ integral of Eqs. (\ref{I1_Wick}, \ref{I1_pole}) with
\ber
g (Q^2) = - \frac{1}{2K^2} \left( \tilde{Q}^2 - \frac{Q^2}{4} \right)^2 \left( \tilde{Q}^2 + \frac{Q^2}{4} \right) \beta \left( \tilde{Q}^2 - \frac{Q^2}{4}\right).
\eer
The integral from the second term of Eq. (\ref{subtraction_3_initial}) can be performed by the Gegenbauer polynomial technique for the angular integration. The result is given by
\ber \label{TPE_subtraction_all_integration}
 \mathop{\mathlarger{\int}} \frac{  i \mathrm{d}^4 \tilde{q}}{\left( 2 \pi \right)^4} \frac{\Pi_Q^+ \Pi_Q^-}{2 K^2} \left( \tilde{Q}^2 + \frac{Q^2}{4} \right) \beta \left( \tilde{Q}^2 - \frac{Q^2}{4}\right) =  \frac{-1}{32 \pi^2 \left( 1 + a \right)} \mathop{\mathlarger{\int}} \limits^{~~\infty}_0 x^n \beta \left( \frac{ \left( x - 1 \right) Q^2}{4} \right) \mathrm{d} x ,
\eer
with $ n = 1 $ for $ x < 1 $ and $ n = 0 $ for $ x > 1 $.

Summing up all contributions we obtain the result of Eqs. (\ref{TPE_subtraction_all_integration_result}) and (\ref{weighting_function_expression}).

\end{document}